\documentclass[preprint,12pt,authoryear]{elsarticle}

\usepackage{amssymb}

\usepackage{amsmath}
\usepackage{hyperref}
\usepackage{lineno}
\usepackage{natbib}
\usepackage{graphicx}
\usepackage{color}
\usepackage{url}
\hypersetup{pdftitle={Plasma distribution of Comet ISON (C/2012 S1) observed using the radio scintillation method}, pdfauthor={Tomoya Iju}}

\newcommand{\degree}{$^\circ$}

\journal{Icarus}

\begin{document}

\begin{frontmatter}



\title{Plasma distribution of Comet ISON (C/2012 S1) observed using the radio scintillation method}


\author[ste]{Tomoya~Iju\corref{cor1}}
\ead{tomoya@stelab.nagoya-u.ac.jp}

\author[nhn]{Shinsuke~Abe}
\ead{avell@aero.cst.nihon-u.ac.jp}

\author[ste]{Munetoshi~Tokumaru}
\ead{tokumaru@stelab.nagoya-u.ac.jp}

\author[ste]{Ken'ichi~Fujiki}
\ead{fujiki@stelab.nagoya-u.ac.jp}

\address[ste]{Solar-Terrestrial Environment Laboratory, Nagoya Univ., Furo-cho, 
Chikusa-ku, Nagoya, Aichi 464-8601, Japan}
\address[nhn]{Dept. of Aerospace Engineering, Coll. Sci. Tech., Nihon Univ., 
7-24-1 Narashinodai, Funabashi, Chiba 274-8501, Japan}
\cortext[cor1]{Corresponding author}

\begin{abstract}
We report the electron density in a plasma tail of Comet ISON (C/2012 S1) derived from interplanetary scintillation 
(IPS) observations during November 1\,--\,28, 2013. Comet ISON showed a well-developed plasma tail 
(longer than $2.98 \times 10^{7}~\mathrm{km}$) before its perihelion passage on November 28. 
We identified a radio source whose line-of-sight approached the ISON's plasma tail in the above period 
and obtained its IPS data using the Solar Wind Imaging Facility at 327 MHz. We used the \textit{Heliospheric Imager} 
onboard the \textit{Solar-Terrestrial Relation Observatory} to distinguish between the cometary tail 
and solar eruption origins of their enhanced scintillation. 
From our examinations, we confirmed three IPS enhancements of a radio source 1148$-$00 on November 13, 16, and 17, 
which could be attributed to the disturbance in the cometary tail. Power spectra of 1148$-$00 had the steeper slope 
than normal ones during its occultation by the plasma tail. We estimated the electron density in the ISON's plasma tail and found 
84 ${\mathrm{cm^{-3}}}$ around the tail axis at a distance of $3.74 \times 10^{7}~\mathrm{km}$ from the cometary nucleus and 
an unexpected variation of the electron density in the vicinity of the tail boundary.
\end{abstract}

\begin{keyword}
Comets \sep Comets, plasma \sep Radio observations \sep Interplanetary medium \sep Solar wind


\end{keyword}

\end{frontmatter}



\section{Introduction}
  \label{introduction}

Comet ISON (C/2012 S1) was found out by Nevski and Novichonok using a 0.4 m telescope of the International Scientific Optical 
Network on September 21, 2012 (\citealp{Nevski2012}). Because it was one of the sun-grazing comets, 
which had a perihelion distance of 0.0125 astronomical units (AU) ($1.87 \times 10^{6}~\mathrm{km}$), Comet ISON was expected to emit 
a large amount of gas and then become extremely bright before and after its perihelion passage. However, 
the ISON's nucleus collapsed on November 28, 2013 when it passed the closest point to the Sun on its orbit 
(\citealp{Knight2014}; \citealp{Lisse2014}), and so far no one has confirmed any on-orbit fragments after the time when Comet ISON's 
remnants went out of the space-borne coronagraph field-of-view (e.g. \url{http://hubblesite.org/hubble_discoveries/comet_ison}). 
During pre-perihelion, Comet ISON showed a well-developed plasma tail. 
The measurement of plasma, particularly its electron density, usually requires an in situ observation by a comet probe. 
Direct plasma measurements have been carried out in the downstream of the cometary nucleus for Comets Giacobini-Zinner 
(\citealp{Meyer-Vernet1986}), Hyakutake (\citealp{Gloeckler2000}), and McNaught (\citealp{Neugebauer2007}). 
However, there was no spacecraft to measure the plasma tail of Comet ISON directly. 

Remote sensing of the cometary plasma tail using radio observations was begun in the 1950s (\citealp{Whitfield1957}). 
\cite{Wright1979} observed some radio source occultation by plasma tails of Comets Kohoutek and West and found peak electron 
densities of approximately 2\,--\,5 $\times 10^{4}$ ${\mathrm{cm^{-3}}}$ in their tails from observed anomalies of radio source positions.
The interplanetary scintillation (IPS) is a phenomenon in which radio signals from distant radio sources fluctuate by density 
irregularities of the solar wind, and it is well known that interplanetary disturbances such as coronal mass ejections (CMEs) 
cause an abrupt increase in IPS (\citealp{Hewish1964}; \citealp{Gapper1982}). A cometary plasma tail may also be a potential cause for 
the IPS enhancement. \cite{Ananthakrishnan1975} observed an IPS of a radio source at 327 MHz during its occultation by 
a plasma tail of Comet Kohoutek. From their result, \cite{Lee1976} estimated the root-mean-square fluctuation of the electron density 
as ${\approx80}$ ${\mathrm{cm^{-3}}}$. After their pioneering work, similar observations have been made for Comets Halley 
(\citealp{Alurkar1986}; \citealp{Ananthakrishnan1987}; \citealp{Slee1987}), Wilson (\citealp{Slee1990}), Austin (\citealp{Janardhan1991}), 
Hale-Bopp (\citealp{Abe1997}), and Schwassmann-Wachmann 3-B (\citealp{Roy2007}). In spite of these studies since 1975, 
the IPS enhancement due to the cometary tail is still controversial. \cite{Alurkar1986} and \cite{Slee1987} presented
positive results, while \cite{Ananthakrishnan1987} reported that no significant enhancement of scintillation was observed for 
a radio source occultation of Comet Halley. Because the IPS observation alone could not distinguish between the cometary tail 
and solar-wind irregularity origins of the enhanced scintillation, it was difficult to obtain a conclusive result for the IPS of the plasma tail. 

In the current study, we examine the IPS enhancement due to the plasma tail of Comet ISON using the radio telescope system
of the Solar-Terrestrial Environment Laboratory (STEL), Nagoya University during November 1\,--\,28, 
2013. To improve limitations of the IPS observation mentioned above, we analyze data of an imaging instrument onboard 
the \textit{Solar-Terrestrial Relation Observatory} (STEREO) spacecraft (\citealp{Kaiser2008}). From these examinations, we estimate 
the electron density in the plasma tail of Comet ISON. 
The outline of this article is as follows: 
Section~\ref{method} describes IPS observations, images taken by STEREO and amateur astronomers, and a method for event identification. 
Section~\ref{results} provides analyses of IPS enhancement events by the ISON's tail. 
Section~\ref{discussion} discusses the results and gives the main conclusion of our study. 

\section{Data and method}
  \label{method}

\subsection{Data}
  
STEL IPS observations at 327 MHz have been carried out regularly using ground-based radio telescopes to investigate 
the solar wind and interplanetary disturbance since the early 1980s (\citealp{Kojima1990}). The Solar Wind Imaging Facility (SWIFT) 
has been in operation since 2010 and capable of observing $\approx40$ radio sources in a day (\citealp{Tokumaru2011}). 
For each radio source, the solar-wind disturbance factor, the so called ``\textit{g}-value'' 
(\citealp{Gapper1982}), is derived from an IPS observation. In this study, we use the \textit{g}-value data obtained using SWIFT. 
A \textit{g}-value represents the relative level of density fluctuation integrated along a line-of-sight from a radio source 
to a radio telescope, and is defined by the following equation 
(\citealp{Tokumaru2003}, \citeyear{Tokumaru2006}; \citealp{Iju2013}): 
\begin{equation} 
  \label{eq:gvalue1} 
g^{2} = \frac{\int_{0}^{\infty}\mathrm{d}z{\{}{\Delta}N_\mathrm{e}(\varepsilon, \psi, r){\}}^{2}{\omega}(z)}{\int_{0}^{\infty} \mathrm{d}z{\{}{\Delta}N_\mathrm{e0}(\varepsilon, \psi, r){\}}^{2}{\omega}(z)}, 
\end{equation} 
where $z$ is the distance along a line-of-sight, $\varepsilon$ and $\psi$ are the heliographic longitude and latitude, respectively, 
$r$ is the radial distance from the Sun, ${{\Delta}N_\mathrm{e}(\varepsilon, \psi, r)}$ is the observed fluctuation level of plasma (electron) density, 
${{\Delta}N_\mathrm{e0}(\varepsilon, \psi, r)}$ is the yearly mean of ${{\Delta}N_\mathrm{e}(\varepsilon, \psi, r)}$, and 
${\omega}(z)$ is the IPS weighting function (\citealp{Young1971}) in a weak scattering condition. 
The ${\omega}(z)$ is given by the following formula (\citealp{Tokumaru2003}):
\begin{equation}
 \label{eq:weightingfunction}
\omega(z) = \int_{0}^{\infty}\mathrm{d}kk^{1-q}\sin^{2}\left(\frac{k^{2}z\lambda}{4\pi}\right)\exp\left(-\frac{k^{2}z^{2}\Theta^{2}}{2}\right),
\end{equation}
where $k$, $q$, $\lambda$, and $\Theta$ are the spatial wavenumber of density fluctuations, the spectral index of the density turbulence, 
the apparent angular size of a radio source, and the wavelength for observing frequency, respectively. We use Equation (\ref{eq:weightingfunction}) 
with $q = 11/3$, $\lambda = 0.92~\mathrm{m}$, and $\Theta = 0.1$'' for STEL IPS observations. 
The ${{\Delta}N_\mathrm{e}(\varepsilon, \psi, r)}$ and ${{\Delta}N_\mathrm{e0}(\varepsilon, \psi, r)}$ are assumed to be proportional 
to the electron density [${N_\mathrm{e}(\varepsilon, \psi, r)}$] and its yearly mean [${N_\mathrm{e0}(\varepsilon, \psi, r)}$], 
respectively (\citealp{Coles1978}). \cite{Ananthakrishnan1980} reported that the relationship between 
${{\Delta}N_\mathrm{e}(\varepsilon, \psi, r)}$ and ${N_\mathrm{e}(\varepsilon, \psi, r)}$ varied with the velocity gradient of the solar wind. 
According to \cite{Asai1998}, however, the relative fluctuation level remained unity within a standard deviation 
in a velocity range of 400\,--\,600 $\mathrm{km~s^{-1}}$ and were dominated mainly by the density rather than 
the velocity of the solar wind for the 327 MHz IPS observation. Therefore, the above assumptions are valid for applying 
to Equation (\ref{eq:gvalue1}), and we obtain: 
\begin{equation} 
  \label{eq:gvalue2} 
g^{2} \propto \frac{\int_{0}^{\infty}\mathrm{d}z{\{}N_\mathrm{e}(\varepsilon, \psi, r){\}}^{2}{\omega}(z)}{\int_{0}^{\infty} \mathrm{d}z{\{}N_\mathrm{e0}(\varepsilon, \psi, r){\}}^{2}{\omega}(z)}. 
\end{equation} 
Because of a normalized index by a mean, the \textit{g}-value is around unity for a quiet condition of the solar wind.
The \textit{g}-value becomes greater than unity with dense plasma or high turbulence on a line-of-sight, 
but lesser than unity for a rarefaction of the solar wind. 

The IPS observation alone cannot distinguish between the cometary plasma tail and CMEs origins of an enhanced \textit{g}-value. 
To confirm CMEs, we use images taken by the \textit{Heliospheric Imager} (HI; \citealp{Eyles2009}) onboard the STEREO-A spacecraft. 
HI comprises the HI-1 and HI-2 cameras, which take images of the interplanetary space between 4\degree~and 24\degree~ 
(elongation from the Sun's center) with a cadence of 40 min and between $18.7$\degree~and $88.7$\degree~with a cadence of two hours, 
respectively. We use their level-2 data processed with a running window of 11 days and differential images, which are available on the STEREO 
Science Center web site (\url{http://stereo-ssc.nascom.nasa.gov}). 

\subsection{Event identification and analysis}
   
We measured the outspread angle and the length of the ISON's plasma tail from photographs taken by two amateur astronomers. 
On November 17, 2013, a narrow-field ($1.87$\degree~$\times~2.89$\degree) image was taken by G. Rhemann in Namibia 
(available at \url{www.astrostudio.at/all.php}), while a wide-field ($8.71$\degree~$\times 5.76$\degree) image was obtained by M. J{\"{a}}ger 
in Austria (available at \url{http://cometpieces-at.webnode.at}). Table \ref{tab:isontail} presents the outspread angle [$\theta_\mathrm{tail}$] and 
length [$L_\mathrm{tail}$] of the ISON's plasma tail derived from these images. These measurements and an ephemeris were used to identify 
radio sources occulted by the plasma tail. 

       %
        \begin{table}[!p]
        \caption{
        Outspread angles and lengths of Comet ISON's plasma tail on November 17, 2013.
        }
        \label{tab:isontail}
        \scalebox{0.90}{
        \begin{tabular}{ccc}
        \hline
~~~ & $\theta_\mathrm{tail}$~(~\degree~) & $L_\mathrm{tail}$~($\times 10^{7}~\mathrm{km}$)  \\
        \hline
Minimum &~4.6~(dense region)~ & 2.98 \\
Maximum &~8.9~(sparse region) & 4.47 \\
        \hline
        \end{tabular}
        }
        \end{table}
     
\begin{figure}[!p]
         \centerline{\includegraphics[width=0.40 \textwidth,clip=]{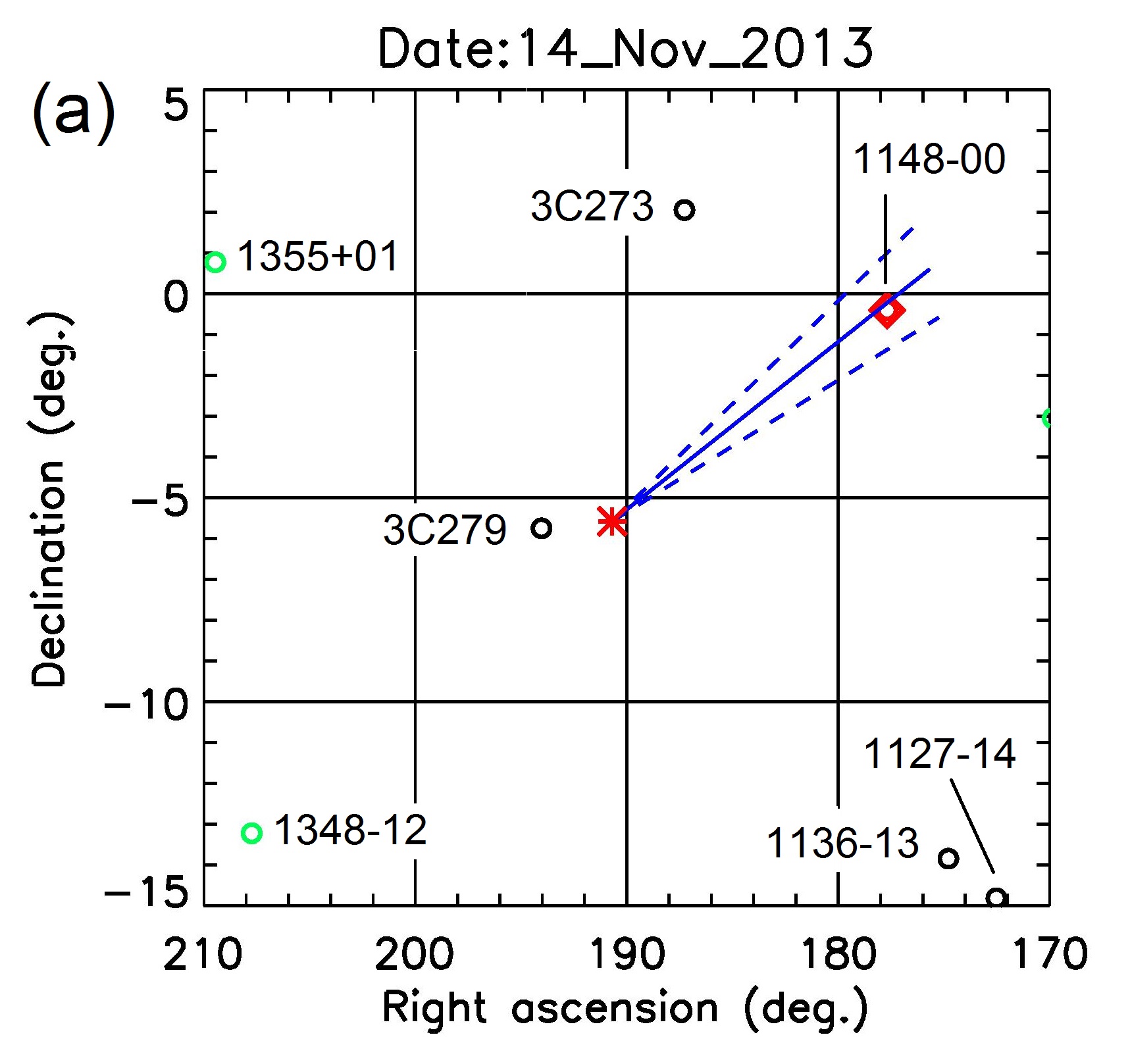}
                     \hspace*{0.01 \textwidth}
                     \includegraphics[width=0.40 \textwidth,clip=]{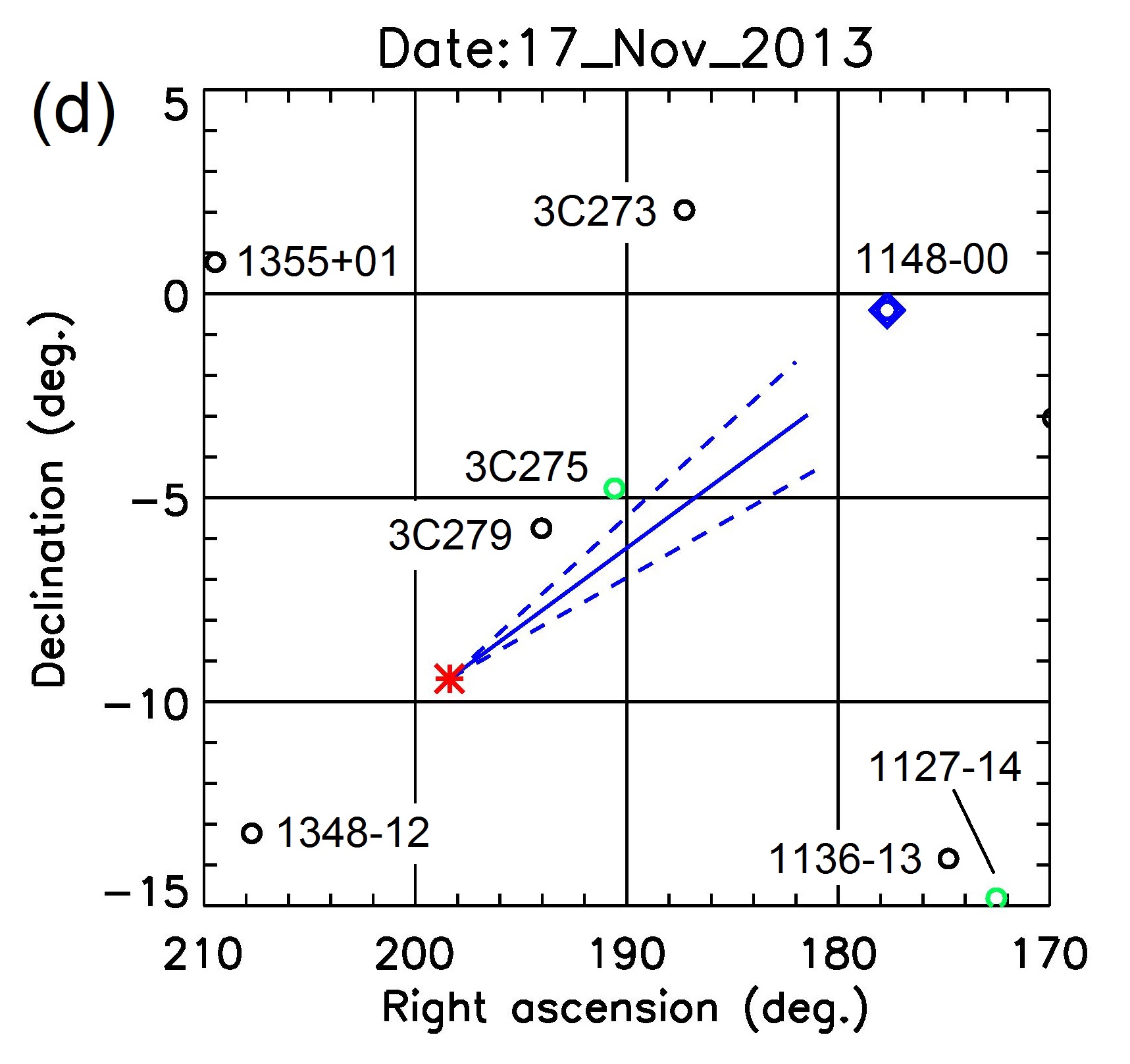}
                     }
           \vspace{0.010 \textwidth} 
         \centerline{\includegraphics[width=0.40 \textwidth,clip=]{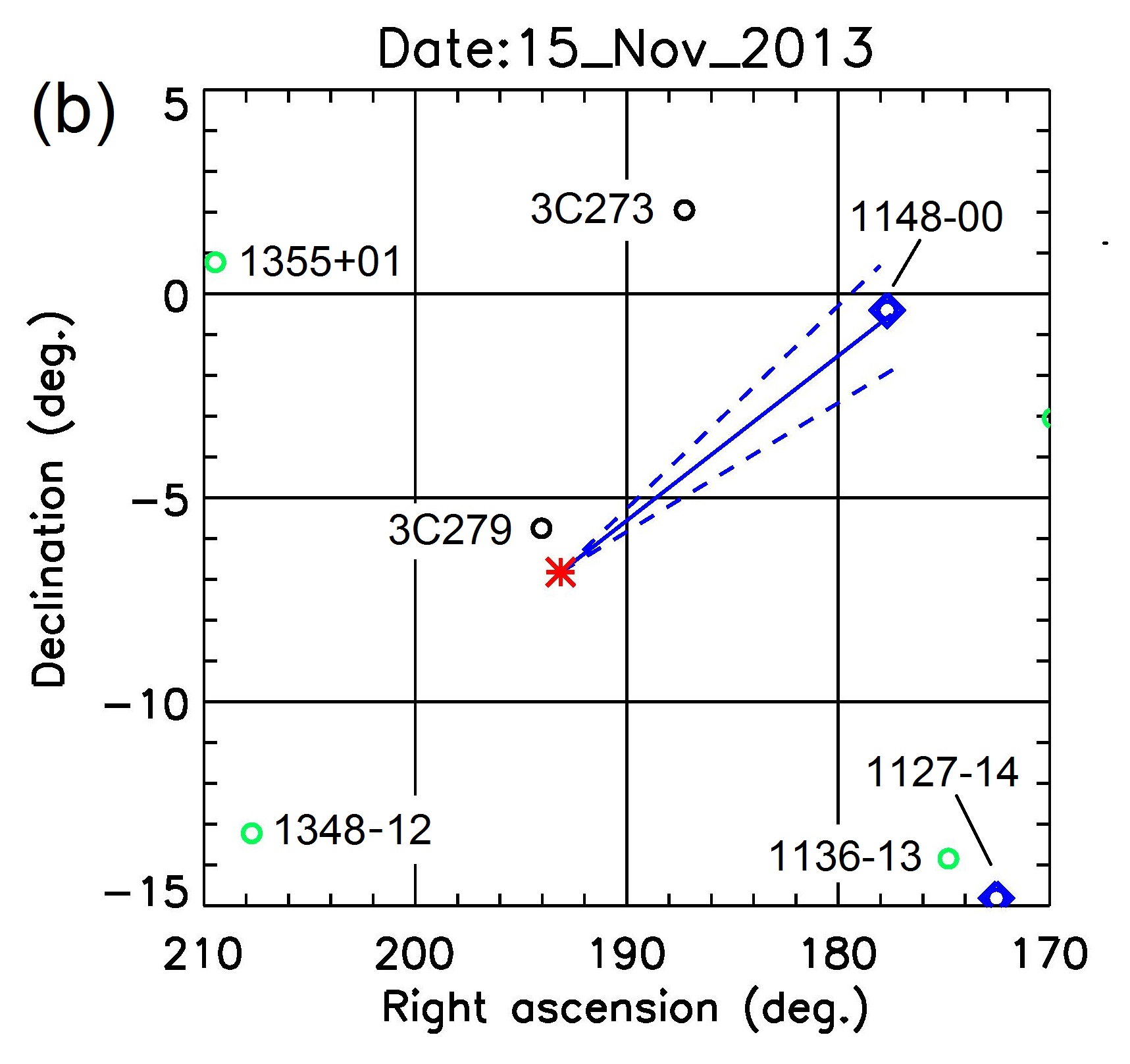}
                     \hspace*{0.01 \textwidth}
                     \includegraphics[width=0.40 \textwidth,clip=]{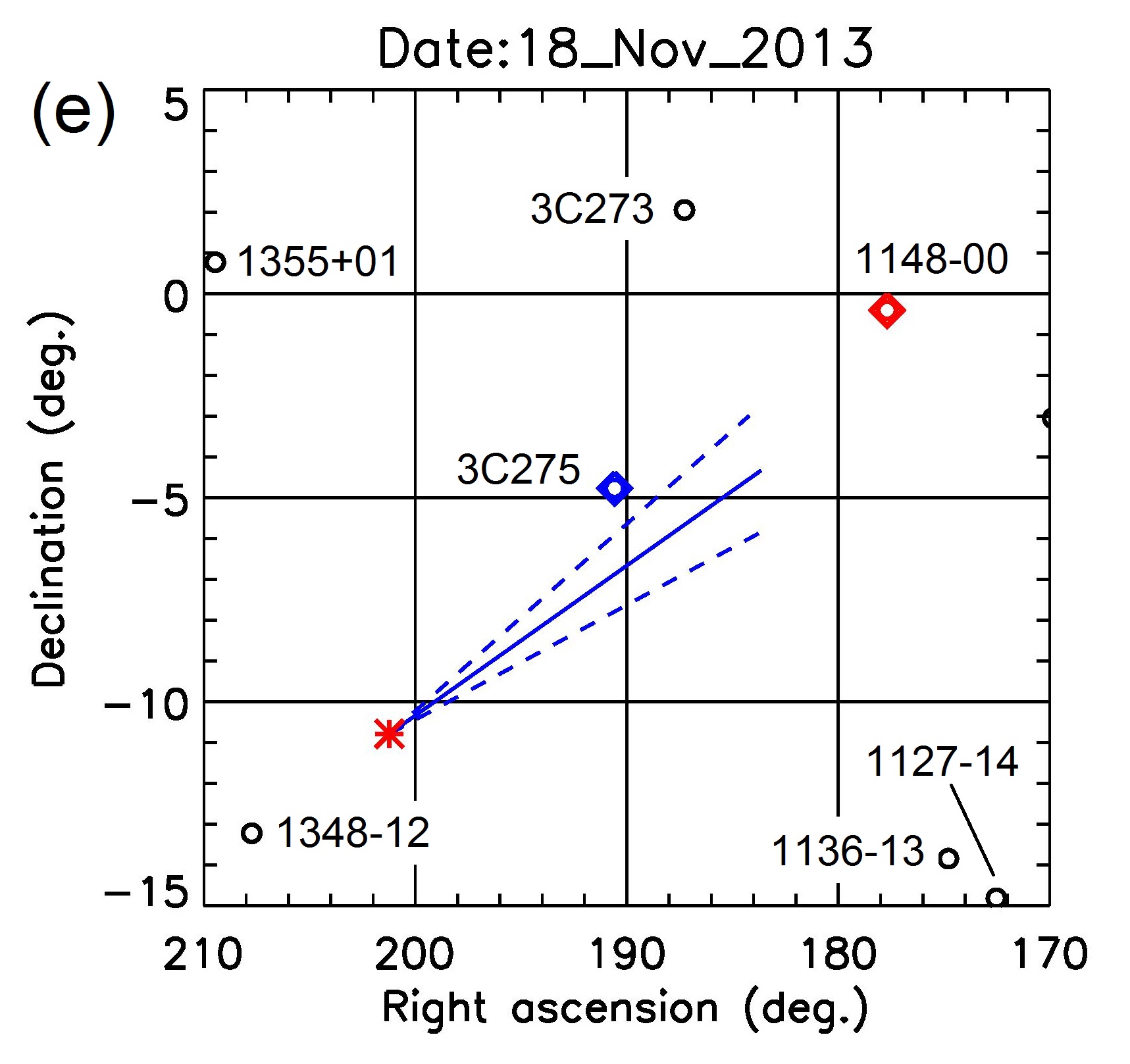}
                     }
           \vspace{0.010 \textwidth} 
         \centerline{\includegraphics[width=0.40 \textwidth,clip=]{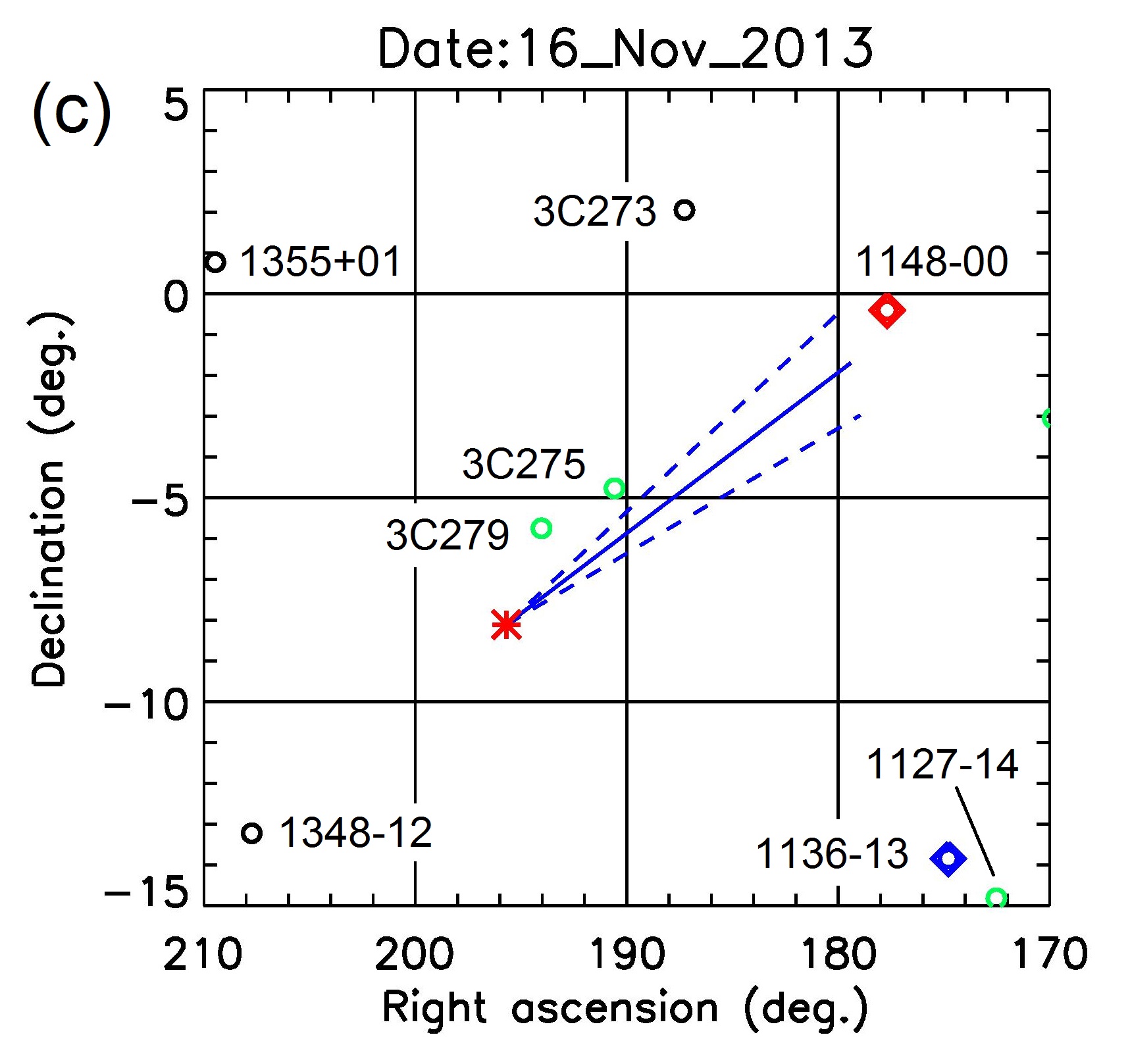}
                     \hspace*{0.01 \textwidth}
                     \includegraphics[width=0.40 \textwidth,clip=]{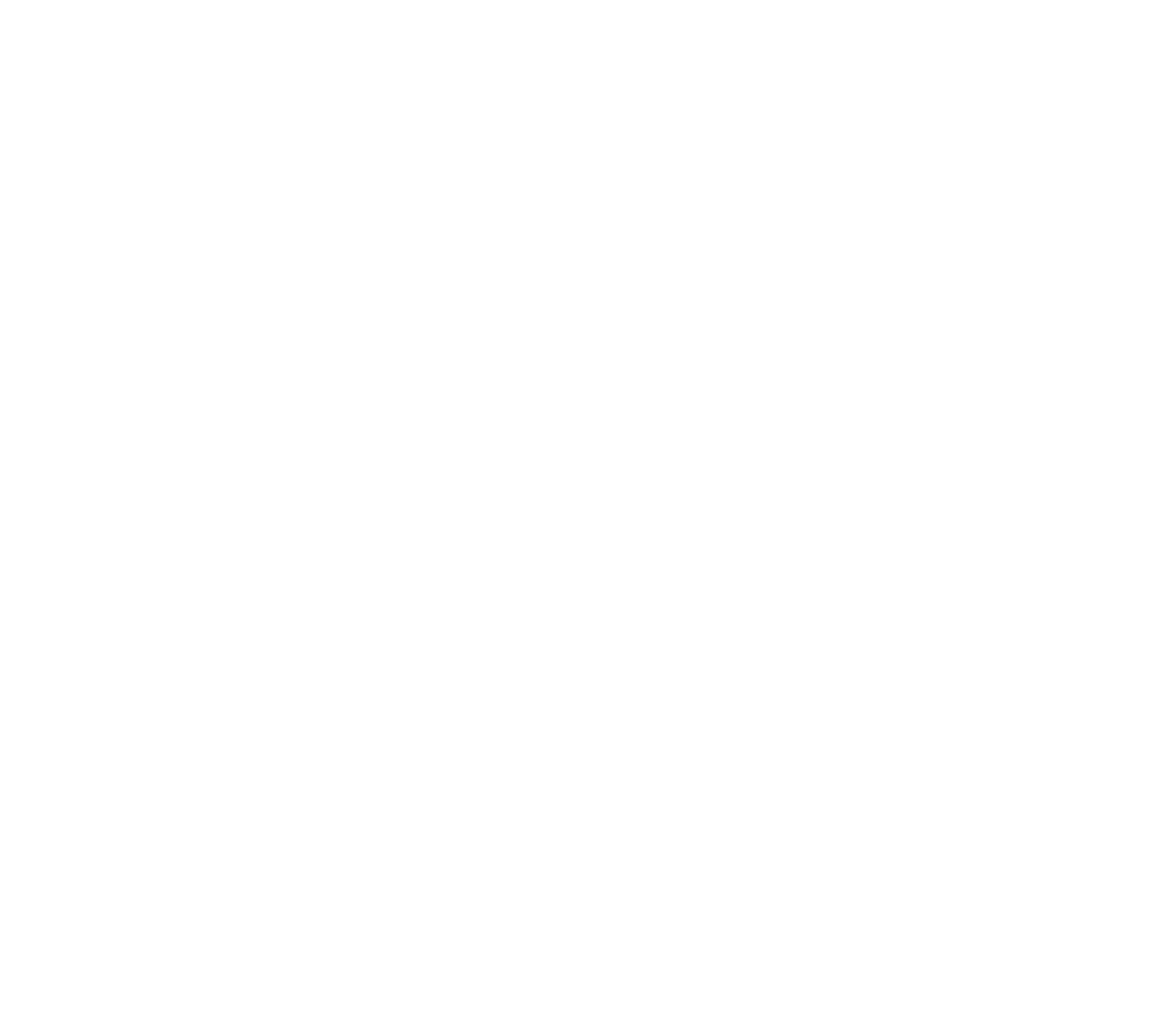}
                     }
           \vspace{0.015 \textwidth} 
         \caption{
         Positions of Comet ISON and nearby radio sources at 00:00 UT on (a) 14, (b) 15, (c) 16, (d) 17, and (e) 18 November
         2013. In each panel, the asterisk (red) and solid line (blue) denote the cometary nucleus and tail axis with a length of 
         $4.47 \times 10^{7}~\mathrm{km}$, respectively. Pairs of broken lines (blue) show an outline of 
         the ISON's plasma tail with $\theta_\mathrm{tail} = 8.9$\degree. Circles and diamonds indicate positions of radio sources with 
         \textit{g}-values $< 1.5$ and $\ge 1.5$, respectively. We use four bins of \textit{g} $< 1.0$ (black), $1.0 \le$ \textit{g} $< 1.5$ 
         (green), $1.5 \le$ \textit{g} $< 2.0$ (blue), and \textit{g} $\ge 2.0$ (red) for the \textit{g}-value in these maps. 
          }
         \label{fig:fig1}
         \end{figure}
\begin{figure}[!p]
         \centerline{\includegraphics[width=1.00 \textwidth,clip=]{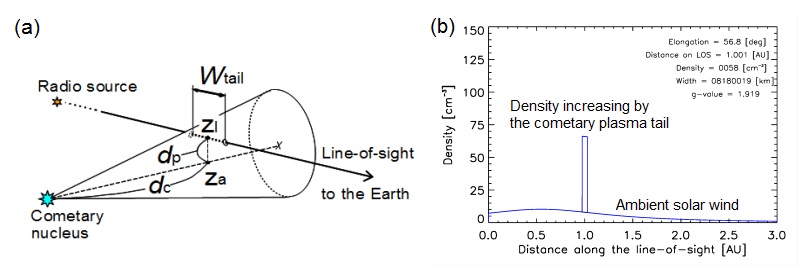}}
           \vspace{-0.010 \textwidth} 
         \caption{
          (a) Schematic explanation of properties for a cone model of the comet system with $8.9$\degree top angle. 
         The ${d_\mathrm{c}}$ is the distance between the cometary nucleus and ${z_\mathrm{a}}$ on the tail axis, 
         ${d_\mathrm{p}}$ is the closest distance between ${z_\mathrm{a}}$ and ${z_\mathrm{l}}$, and ${W_\mathrm{tail}}$ 
         is the width of the ISON's plasma tail intersected by the line-of-sight. 
         (b) Distribution of the electron density along the line-of-sight from 1148$-$00 for the 17 November 2013 IPS enhancement. 
         A rectangular density enhancement with ${W_\mathrm{tail}}$ is given by Comet ISON's plasma tail as the unique perturbation source. 
         }
         \label{fig:fig2}
         \end{figure}

We obtained \textit{g}-values of $\approx40$ radio sources and made a map of them in the sky plane, the so-called ``\textit{g}-map'' 
(\citealp{Gapper1982}), for each day during November 1\,--\,28, 2013. A location of Comet ISON with an outline of the plasma tail was examined 
with respect to radio sources on a \textit{g}-map for each day. From this examination, we found that the radio source 1148$-$00 
(R.A. = $11^\mathrm{h}~50^\mathrm{m}~46^\mathrm{s}$, Dec. = $-$00\degree~24'~13'' in the J2000.0 coordinates) was occulted by the ISON's tail 
during the 12\,--\,18th. We assumed that a \textit{g}-value of 1.5 or more indicates a disturbance of plasma (\citealp{Iju2013}), 
and identified then five candidates for the IPS enhancement by the cometary tail. 
These scintillation enhancements were observed on 1148$-$00 between the 13th at 23:09 and the 17th 
at 22:53 UT; we gained a \textit{g}-value $= 1.375$ on the same source at 23:13 UT on the 12th. 
On the other hand, we identified other sources without the occultation as controls. Table \ref{tab:cometison} summarizes 
an ephemeris of Comet ISON with the plasma tail from November 14 until 18. Figure \ref{fig:fig1} shows a daily change of Comet ISON's position 
with respect to radio sources including 1148$-$00 in the same period. From these observations, we estimated the number density of electrons 
in the plasma tail. 
For each IPS enhancement event, we calculated the intersection point [${z_\mathrm{a}}$] of the cometary tail axis and the projected line-of-sight on 
a tail-axial surface, the distance [${d_\mathrm{c}}$] between the cometary nucleus and ${z_\mathrm{a}}$, and the closest distance [${d_\mathrm{p}}$] 
between ${z_\mathrm{a}}$ and the line-of-sight from 1148$-$00. The ${z_\mathrm{l}}$ was defined as the closest point to ${z_\mathrm{a}}$ on 
the line-of-sight. It is noted that the distance between the cometary nucleus and ${z_\mathrm{l}}$ is given by $(d_\mathrm{c}^{2} + d_\mathrm{p}^{2})^{1/2}$. 
Using ${d_\mathrm{c}}$ and ${d_\mathrm{p}}$, we calculated the thickness of the plasma tail [$W_\mathrm{tail}$] which intersected 
by the line-of-sight with a cone model of the comet system with $8.9$\degree~ of the top angle. These properties are explained schematically in the panel (a) 
in Figure \ref{fig:fig2}. 
We assumed that the density of the solar wind varied with $r$ only, namely an isotropic flow of the solar wind, 
although its actual distribution was more complex along the line-of-sight. 
The inverse-square law of the radial distance was expressed by: 
\begin{equation}
  \label{eq:function}
\kappa(z, \phi) = \frac{1}{1 + z^{2} - 2z\cos\phi},
\end{equation}
where $\phi$ was the solar elongation of a radio source, along its line-of-sight in AU. 
The first term of a solar-wind density model (\citealp{Leblanc1998}) was used as the unperturbed solar wind, i.e. 
$N_\mathrm{e0} = 7.2$ at $r = 1$ AU ($z = 0$ AU). 
A density enhancement region of the width ${W_\mathrm{tail}}$ by the cometary tail was assumed 
as the perturbation source at ${z_\mathrm{l}}$ on the line-of-sight for the perturbed solar wind. 
The uniformity of density was also assumed in this region. 
These assumptions were expressed as the following equation: 
\begin{equation} 
  \label{eq:cases} 
N_\mathrm{e}\kappa(z, \phi) = 
\begin{cases} 
7.2\kappa(z, \phi) & (z_\mathrm{l} - W_\mathrm{tail}/2 > z~\mathrm{or}~z_\mathrm{l} + W_\mathrm{tail}/2 < z), \\ 
7.2\kappa(z, \phi) + n_\mathrm{e} & (z_\mathrm{l} - W_\mathrm{tail}/2 \le z \le z_\mathrm{l} + W_\mathrm{tail}/2), 
\end{cases} 
\end{equation} 
where ${n_\mathrm{e}}$ was the average density of electrons. For an IPS enhancement, 
an example of the distribution of electron density is shown in the panel (b) in Figure \ref{fig:fig2}. 
We integrated Equation (\ref{eq:gvalue2}) numerically with Equation (\ref{eq:cases}) and a given $\phi$ to results in the observed \textit{g}-value 
and then obtained the value of ${n_\mathrm{e}}$ in the ISON's plasma tail for each IPS event. 
        
       %
        \begin{table}[!p]
        \caption{
        Ephemeris of Comet ISON with the plasma tail having a length of $4.47 \times 10^{7}~\mathrm{km}$.
        }
        \label{tab:cometison}
        \scalebox{0.85}{
        \begin{tabular}{ccccccccc}
        \hline
Date at &  \multicolumn{5}{c}{Nucleus} & & \multicolumn{2}{c}{Plasma tail axis} \\ \cline{2-6} \cline{8-9} 
00:00 UT & \multicolumn{2}{c}{J2000.0 coordinates} & Heliocentric & Geocentric & Elong. & & \multicolumn{2}{c}{Tip deviation from the nucleus} \\
~~~ & R.A. & Dec. & distance & distance & ~~~ & & $\delta$R.A. & $\delta$Dec. \\
~~~ & (h m) & (~\degree~'~) & (AU) & (AU) & (~\degree~) & & (~'~) & (~'~) \\
        \hline
Nov.~14 & 12~42.79 & ~$-$5~34.4 & 0.651 & 0.925 & 39 & & ~$-$900.8 & 370.5 \\
Nov.~15 & 12~52.52 & ~$-$6~49.1 & 0.621 & 0.909 & 37 & & ~$-$938.5 & 378.7 \\
Nov.~16 & 13~02.77 & ~$-$8~06.4 & 0.590 & 0.894 & 36 & & ~$-$976.9 & 384.7 \\
Nov.~17 & 13~13.55 & ~$-$9~25.8 & 0.558 & 0.882 & 34 & & $-$1015.6 & 387.7 \\
Nov.~18 & 13~24.90 & $-$10~47.0 & 0.525 & 0.872 & 32 & & $-$1054.0 & 387.1 \\
        \hline
        \end{tabular}
        }
        \end{table}
     

        %

        %
        
        %
         \begin{figure}[!p]
         \centerline{\includegraphics[width=0.50 \textwidth,clip=]{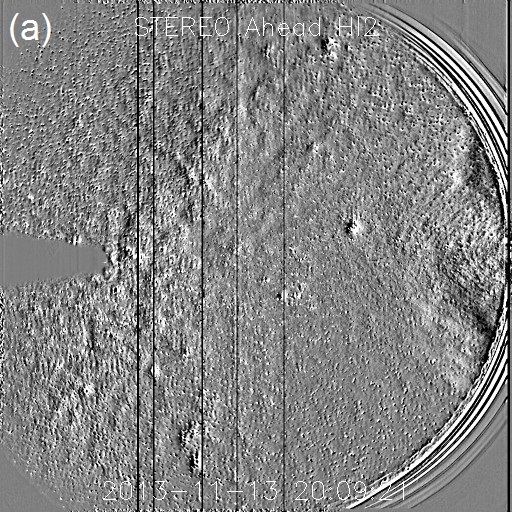}
                     \hspace*{0.01 \textwidth}
                     \includegraphics[width=0.50 \textwidth,clip=]{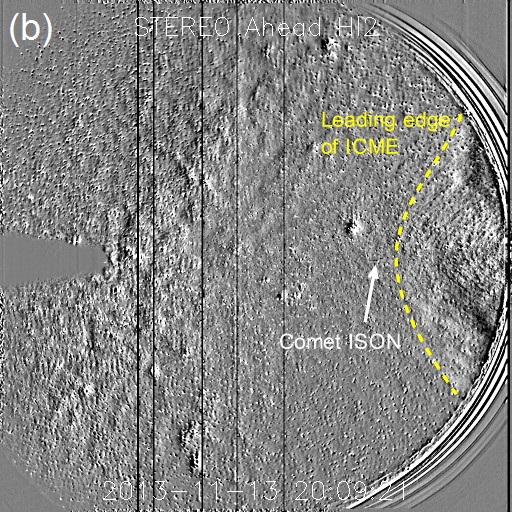}
                     }
           \vspace{-0.010 \textwidth} 
         \caption{
         (a) White-light difference image taken by the STEREO-A/HI-2 for an ICME with Comet ISON at 20:09 UT on November 13, 2013 
         and (b) that with indications of the ICME's leading edge and Comet ISON. 
         }
         \label{fig:fig3}
         \end{figure}

For the above IPS enhancements of 1148$-$00, we examined whether CMEs passed across 
the line-of-sight from that using data of the STEREO-A/HI. 
The STEREO-A was $\approx149$\degree~ahead of the Earth with a heliocentric distance of 0.96 AU between November 13 and 18. 
For interplanetary transients during the 13\,--\,18th, we made HI-2 images into a movie, and checked up CMEs around Comet ISON from 
the movie by eyes because their transient signatures were too faint 
to identify their leading and trailing edges by a brightness measurement. 
From this examination, we confirmed that an interplanetary CME (ICME) passed through Comet ISON from November 14 at 00:00 UT until 16 at the same. 
Figure \ref{fig:fig3} shows a HI-2 image of the ICME and Comet ISON at 20:09 UT on the 13th. 
This ICME was observed as a CME with a velocity of $548~\mathrm{km~s^{-1}}$ by the \textit{Large Angle and Spectrometric Coronagraph} 
(LASCO; \citealp{Brueckner1995}) onboard the \textit{Solar and Heliospheric Observatory} (SOHO) at 06:48 UT, November 12, which was listed in 
the SOHO/LASCO CME catalog (\citealp{Yashiro2004}; available at \url{http://cdaw.gsfc.nasa.gov/CME_list/index.html}). 
Therefore, we considered the same ICME crossed the line-of-sight during the above period, and both the cometary plasma tail and ICME 
contributed to IPS enhancements of 1148$-$00 on the 14th and 15th. 
As a result, we identified three IPS enhancement events on November 13, 16, and 17, which are probably caused by the ISON's tail only. 

\section{Results}
  \label{results}

\begin{figure}[!p]
         \centerline{\includegraphics[width=0.40 \textwidth,clip=]{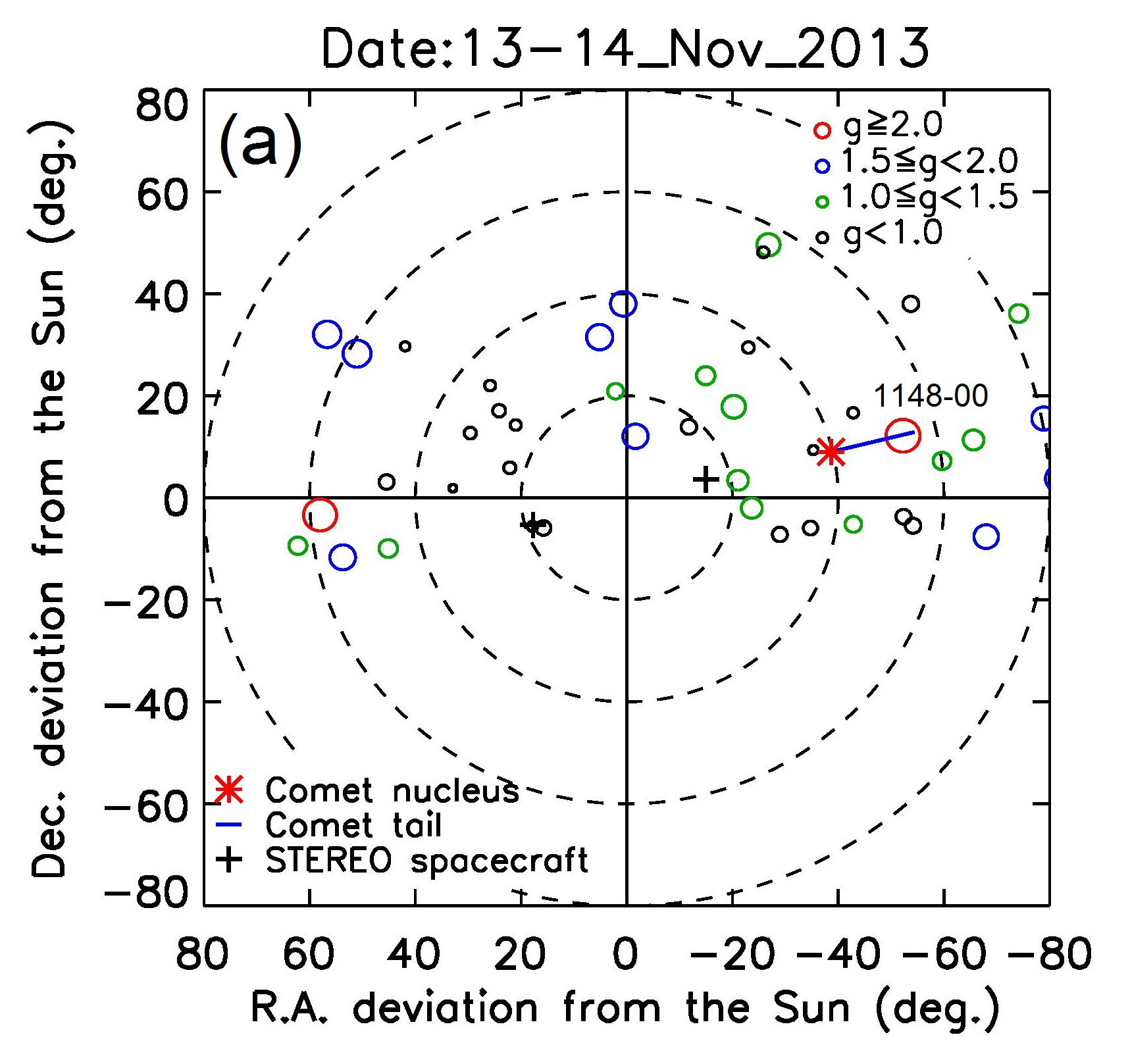}
                     \hspace*{0.01 \textwidth}
                     \includegraphics[width=0.40 \textwidth,clip=]{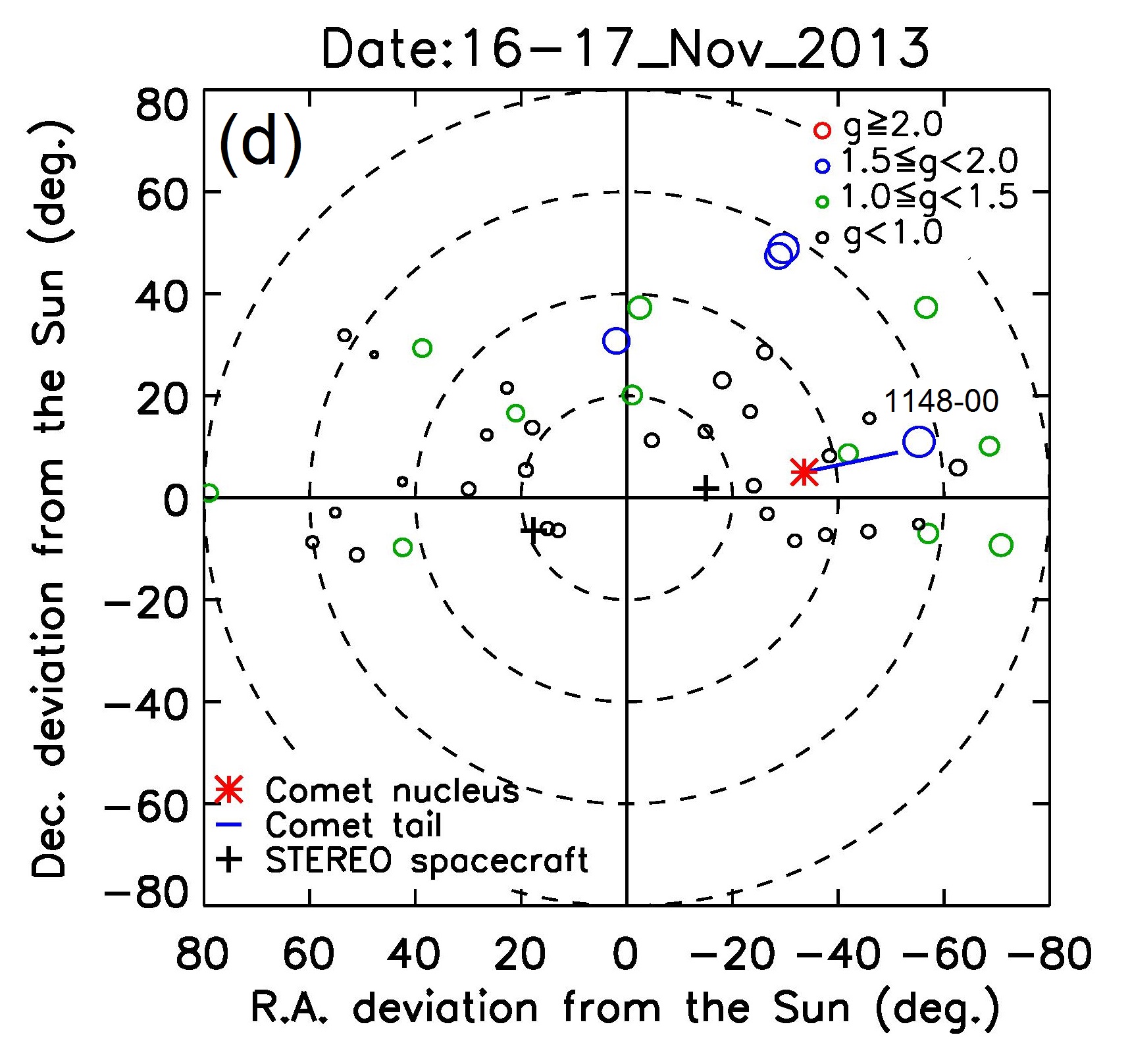}
                     }
           \vspace{0.000 \textwidth} 
         \centerline{\includegraphics[width=0.40 \textwidth,clip=]{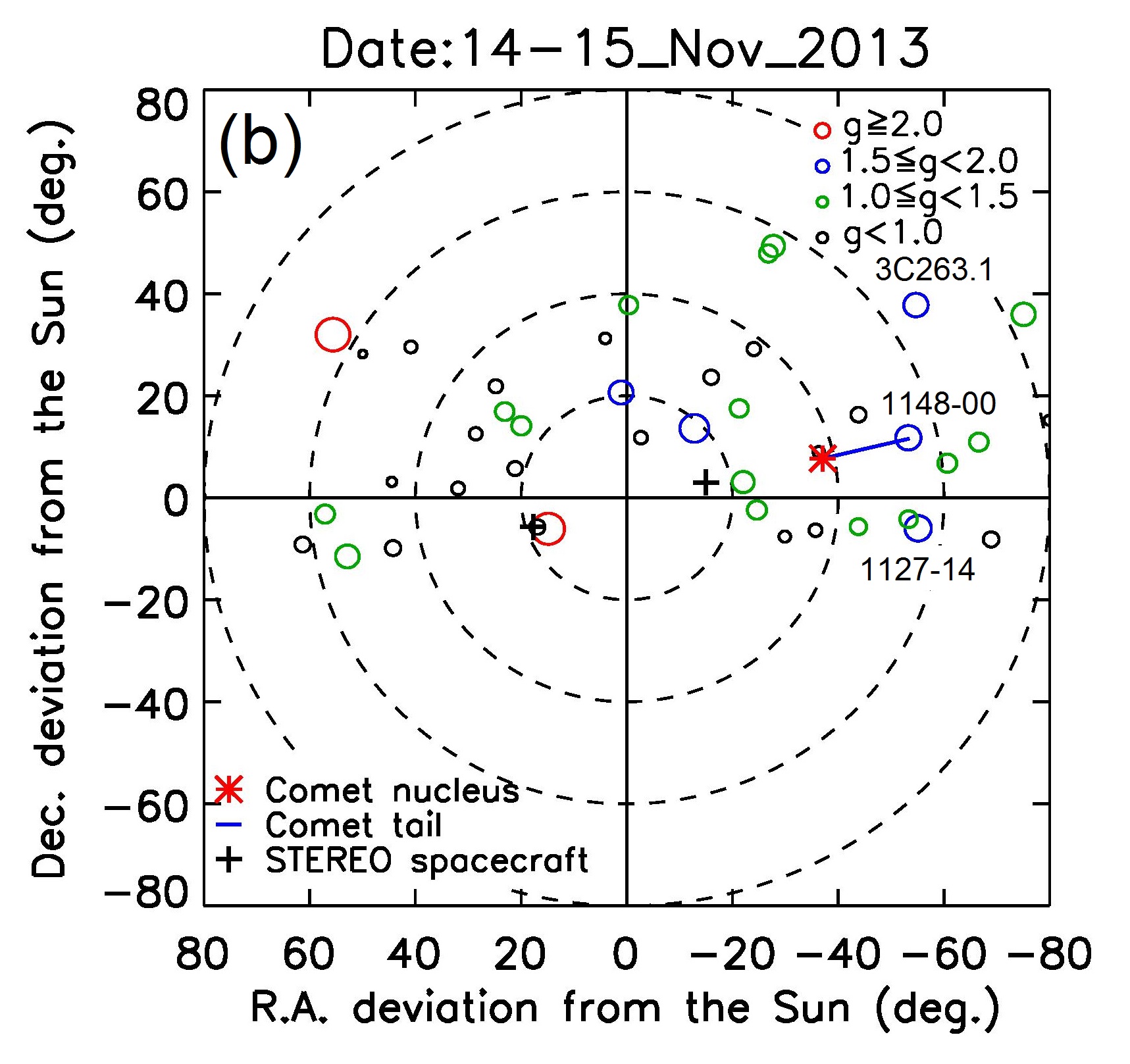}
                     \hspace*{0.01 \textwidth}
                     \includegraphics[width=0.40 \textwidth,clip=]{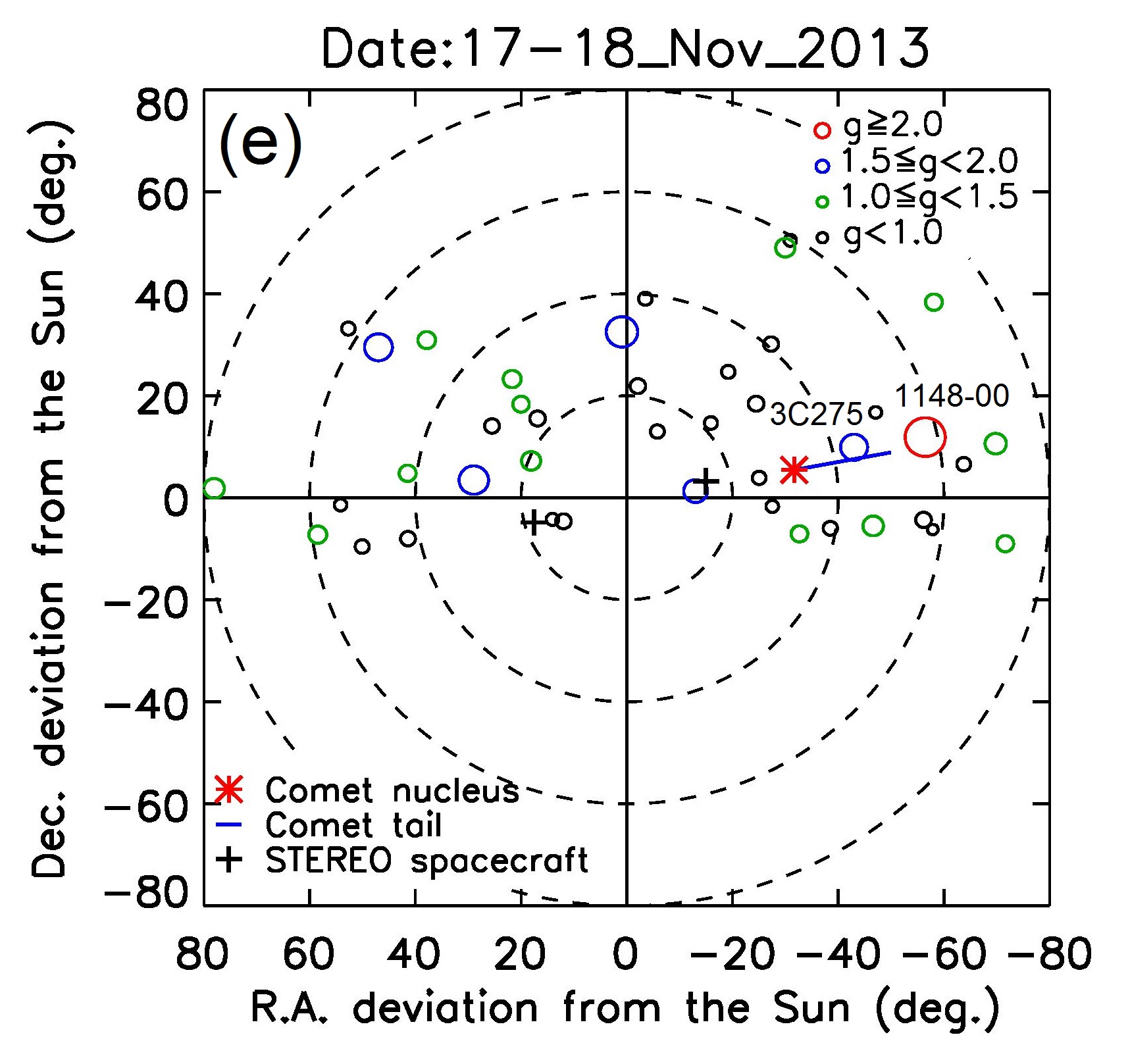}
                     }
           \vspace{0.000 \textwidth} 
         \centerline{\includegraphics[width=0.40 \textwidth,clip=]{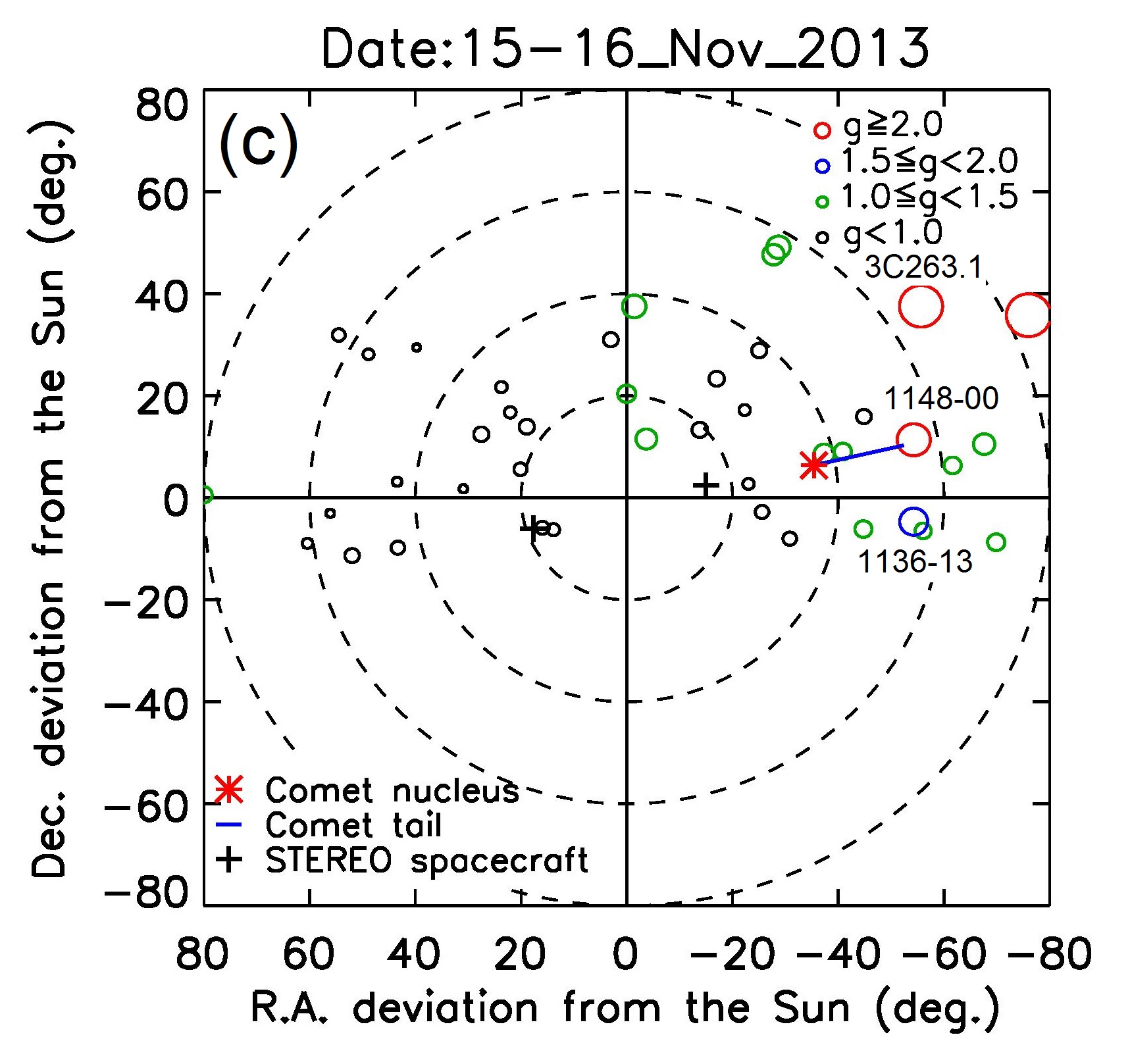}
                     \hspace*{0.01 \textwidth}
                     \includegraphics[width=0.40 \textwidth,clip=]{blank.jpg}
                     }
           \vspace{-0.006 \textwidth} 
         \caption{
         The \textit{g}-maps with Comet ISON during (a) 13 at 21:11\,--\,14 at 08:04, (b) 14 at 20:58\,--\,15 at 08:00, 
         (c) 15 at 20:55\,--\,16 at 07:56, (d) 16 at 20:50\,--\,17 at 07:52, and (e) 17 at 20:47\,--\,18 at 07:48, November 
         2013 in UT. In each panel, the center of a map corresponds to the location of the Sun, and concentric circles indicate 
         the solar elongations of $20$\degree, $40$\degree, $60$\degree, and $80$\degree. 
         Comet ISON is depicted for the nucleus (asterisk) and tail axis (solid line) at 00:00 in each observational period. 
         Crosses denote locations of the STEREO-A and -B spacecraft.
         Colored circles indicate the positions of radio sources and a diameter and color of the circles denote a 
         \textit{g}-value for the sources. Four bins of \textit{g} $< 1.0$ (black), $1.0 \le$ \textit{g} $< 1.5$ 
         (green), $1.5 \le$ \textit{g} $< 2.0$ (blue), and \textit{g} $\ge 2.0$ (red) are used for the \textit{g}-map. 
         Radio sources with an enhanced $g$-value listed in Tables \ref{tab:radiosources1} and \ref{tab:radiosources2} 
         are designated by their name.
          }
         \label{fig:fig4}
         \end{figure}

The \textit{g}-values of the occulted source 1148$-$00 and six controls observed during the 13\,--\,18th are listed in Tables \ref{tab:radiosources1} 
and \ref{tab:radiosources2} with their observation time and elongation. Figure \ref{fig:fig4} shows \textit{g}-maps with Comet ISON, which depict not 
only radio sources in these tables but also other controls. We obtained \textit{g}-values with a time interval of 23 h and 55 min 
for each radio source because SWIFT was not steerable and observed an IPS of radio sources around their local meridian transit. 
We note that radio sources were scanned by SWIFT from the west to east in the sky plane with the Earth's rotation to make a $g$-map in a day, 
and so they could not be observed simultaneously. From Tables  \ref{tab:radiosources1}, \ref{tab:radiosources2}, and 
Figure \ref{fig:fig4}, we find that only 1148$-$00 exhibits \textit{g}-value larger than 1.5 for radio sources being close by Comet ISON on November 
13 and 16, while three sources including that show such high \textit{g}-values on 14 and 15. In panels (b) and (c) of Figure \ref{fig:fig4}, 
radio sources 1127$-$14, 1136$-$13, and 3C263.1 are shown as blue and red circles nearby the 60\degree~curve on the west 
side, which probably relate to the ICME identified by the STEREO-A/HI-2. 

        %

       %
        \begin{table}[!p]
        \caption{
        Positions of occulted and control radio sources and their \textit{g}-values at 327 MHz on 13 and 14 November 2013.
        }
        \label{tab:radiosources1}
        \scalebox{0.90}{
        \begin{tabular}{lccccccccc}
        \hline
Source & \multicolumn{2}{c}{J2000.0 coordinates} & \multicolumn{3}{c}{Nov. 13} &~& \multicolumn{3}{c}{Nov. 14} \\  
\cline{4-6} \cline{8-10} 
~~~ & R.A. & Dec. & Time & $\phi$ & \textit{g}-value & & Time & $\phi$ & \textit{g}-value \\
~~~ & (h~m~s) & (~\degree~'~''~) & (UT) & (~\degree~) & ~~~ & & (UT) & (~\degree~) & ~~~  \\
        \hline
1127$-$14~ & 11~30~05 & $-$14~48~47 & 22:49 & 54 & 0.980 & &  22:44 & 55 & 1.665 \\
1136$-$13~ & 11~39~10 & $-$13~50~34 & 22:58 & 52 & 0.957 & &  22:53 & 53 & 1.099 \\
3C263.1~~ & 11~43~26 & $+$22~08~20 & 23:02 & 66 & 1.000 & &  22:58 & 67 & 1.536 \\
1148$-$00\textsuperscript{a} & 11~50~45 & $-$00~24~13 & 23:09 & 54 & 2.139 & &  23:05 & 55 & 1.594 \\
1213$-$17~ & 12~15~46 & $-$17~31~44 & 23:34 & 43 & 1.042 & &  23:30 & 44 & 1.037 \\
3C273~~ & 12~29~06 & $+$02~03~12 & 23:47 & 46 & 0.682 & &  23:43 & 47 & 0.961 \\
3C275\textsuperscript{A}~~ & 12~42~19 & $-$00~46~02 & --- & --- & --- & & --- & --- & --- \\
        \hline
\multicolumn{10}{l}{
\textsuperscript{A} ``---'' means no observation on Nov. 13 and 14.
} \\
\multicolumn{10}{l}{
\textsuperscript{a} Occulted source
} \\
        \end{tabular}
        }
        \end{table}

       %
        \begin{table}[!p]
        \caption{
        The \textit{g}-values of occulted and control sources at 327 MHz on 15, 16 and 17 November 2013.
        }
        \label{tab:radiosources2}
        \scalebox{0.90}{
        \begin{tabular}{lccccccccccc}
        \hline
Source\textsuperscript{b} & \multicolumn{3}{c}{Nov. 15} & & \multicolumn{3}{c}{Nov. 16} & & \multicolumn{3}{c}{Nov. 17} \\  
\cline{2-4} \cline{6-8} \cline{10-12}
~~~ & Time & $\phi$ & \textit{g}-value & & Time & $\phi$ & \textit{g}-value & & Time & $\phi$ & \textit{g}-value \\
~~~ & (UT) & (~\degree~) & ~~~ & & (UT) & (~\degree~) & ~~~ & & (UT) & (~\degree~) & ~~~ \\
        \hline
1127$-$14~~~ & 22:41 & 56 & 1.017 & & 22:37 & 57 & 1.200 & & 22:33 & 58 & 0.674 \\
1136$-$13~~~ & 22:50 & 54 & 1.769 & & 22:46 & 55 & 0.662 & & 22:42 & 56 & 0.995 \\
3C263.1~~~~ & 22:54 & 68 & 2.717 & & 22:50 & 68 & 1.315 & & 22:46 & 70 & 1.067 \\
1148$-$00\textsuperscript{a} & 23:01 & 56 & 2.093 & & 22:57 & 56 & 1.913 & & 22:53 & 57 & 2.527 \\
1213$-$17~~~ & 23:26 & 45 & 1.078 & & 23:22 & 46 & 0.835 & & 23:19 & 47 & 1.290 \\
3C273~~~~~ & 23:40 & 48 & 0.931 & & 23:35 & 49 & 0.693 & & 23:32 & 50 & 0.759 \\
3C275~~~~~ & 23:53 & 42 & 1.086 & & 23:49 & 43 & 1.159 & & 23:45 & 44 & 1.699 \\
        \hline
\multicolumn{12}{l}{
\textsuperscript{a} Occulted source
} \\
\multicolumn{12}{l}{
\textsuperscript{b} Identical with the column 1 in Table \ref{tab:radiosources1}.
} \\
        \end{tabular}
        }
        \end{table}

\begin{figure}[!p]
      \begin{center}
         \centerline{\includegraphics[width=0.50 \textwidth,clip=]{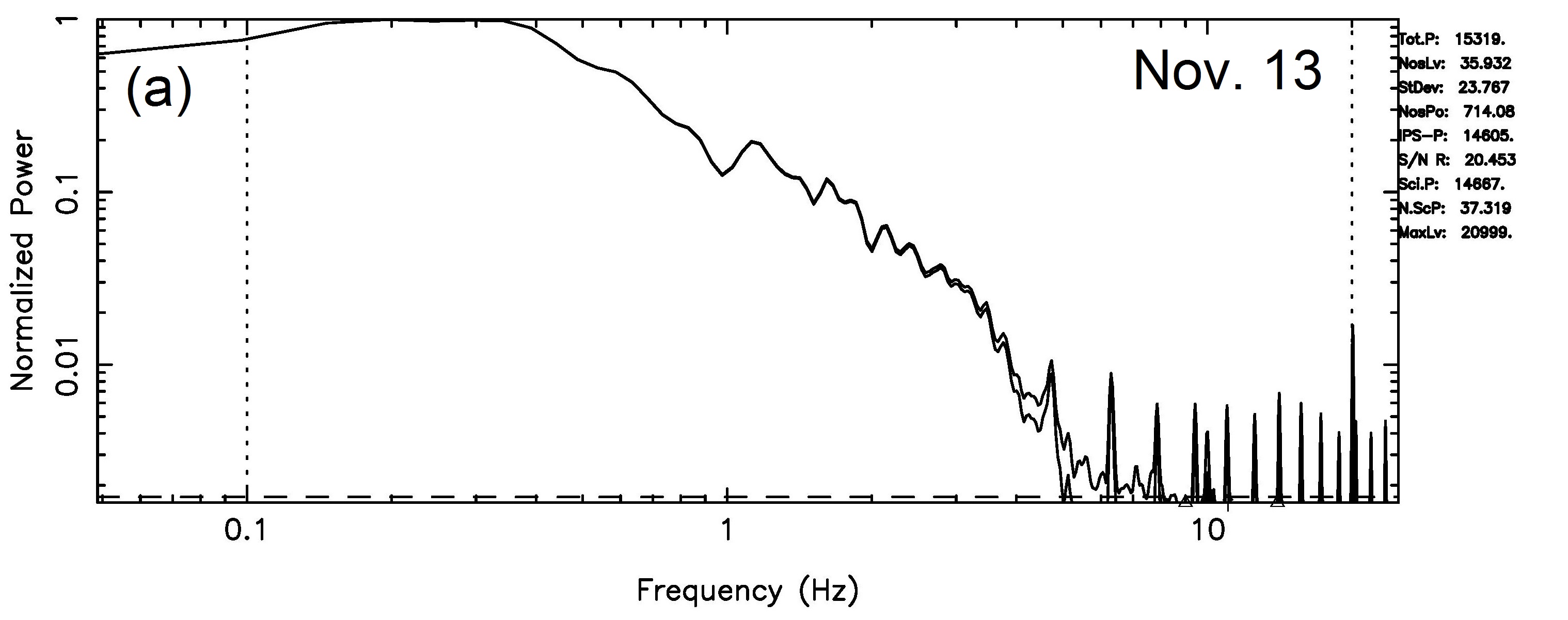}
                     \hspace*{0.00 \textwidth}
                     \includegraphics[width=0.50 \textwidth,clip=]{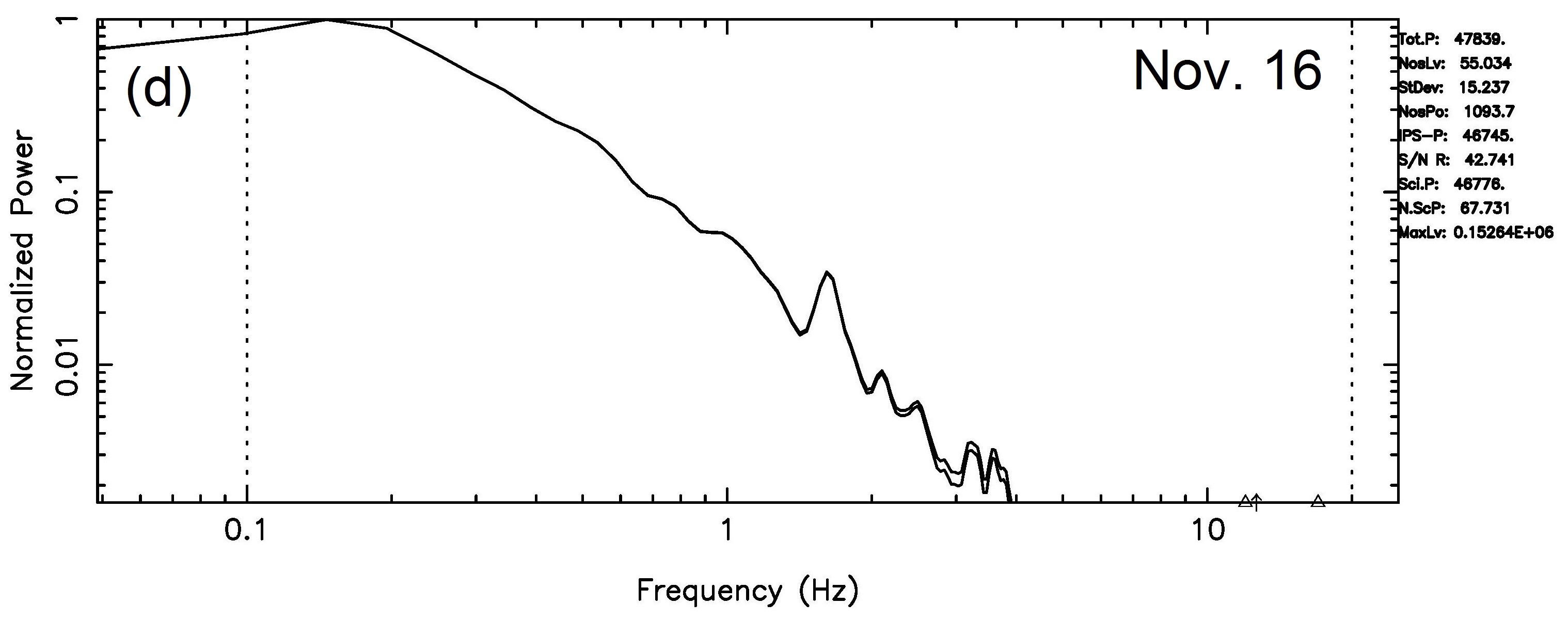}
                     }
           \vspace{0.010 \textwidth} 
         \centerline{\includegraphics[width=0.50 \textwidth,clip=]{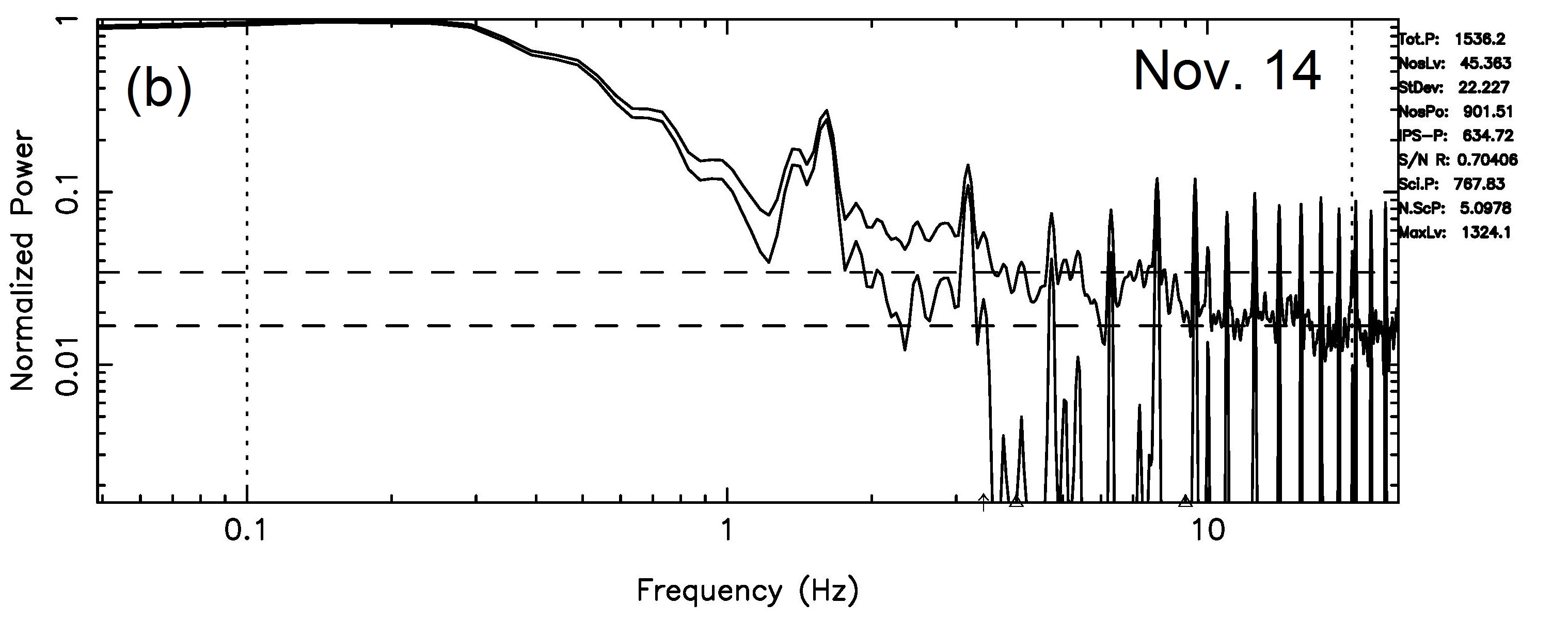}
                     \hspace*{0.00 \textwidth}
                     \includegraphics[width=0.50 \textwidth,clip=]{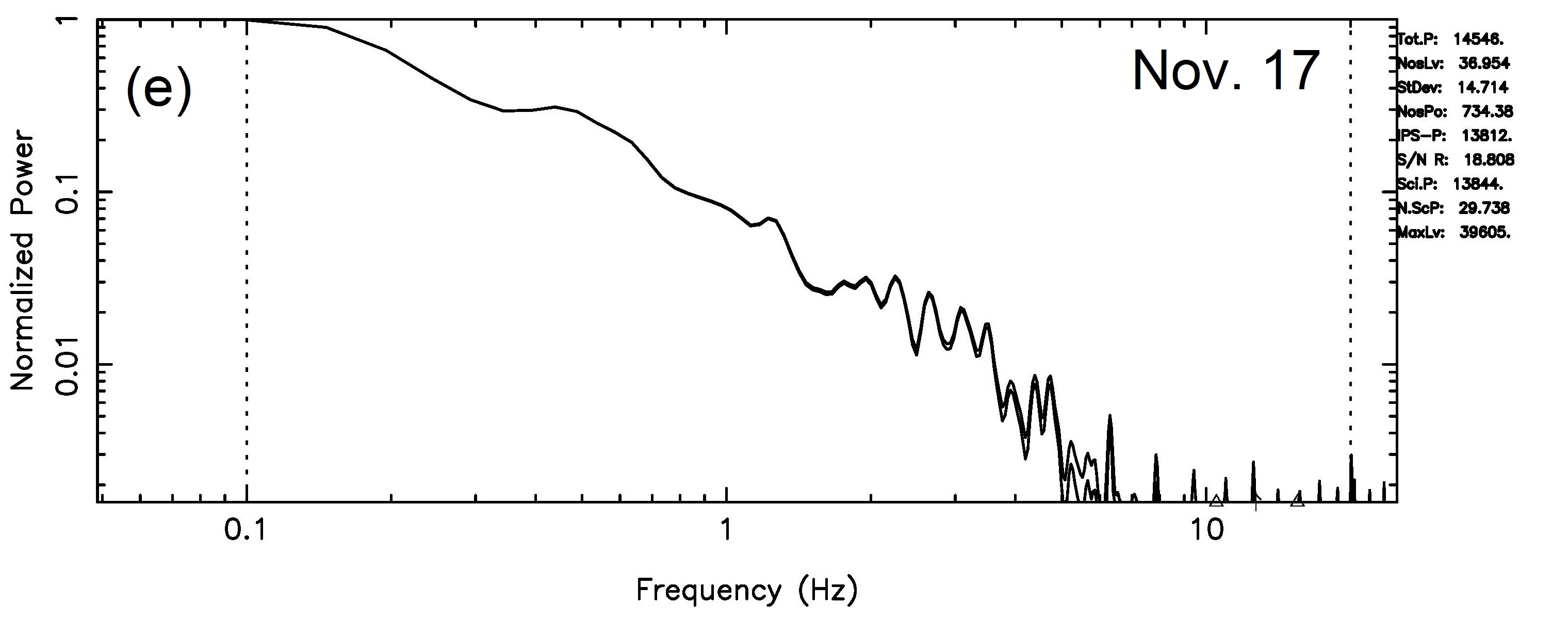}
                     }
           \vspace{0.010 \textwidth} 
         \centerline{\includegraphics[width=0.50 \textwidth,clip=]{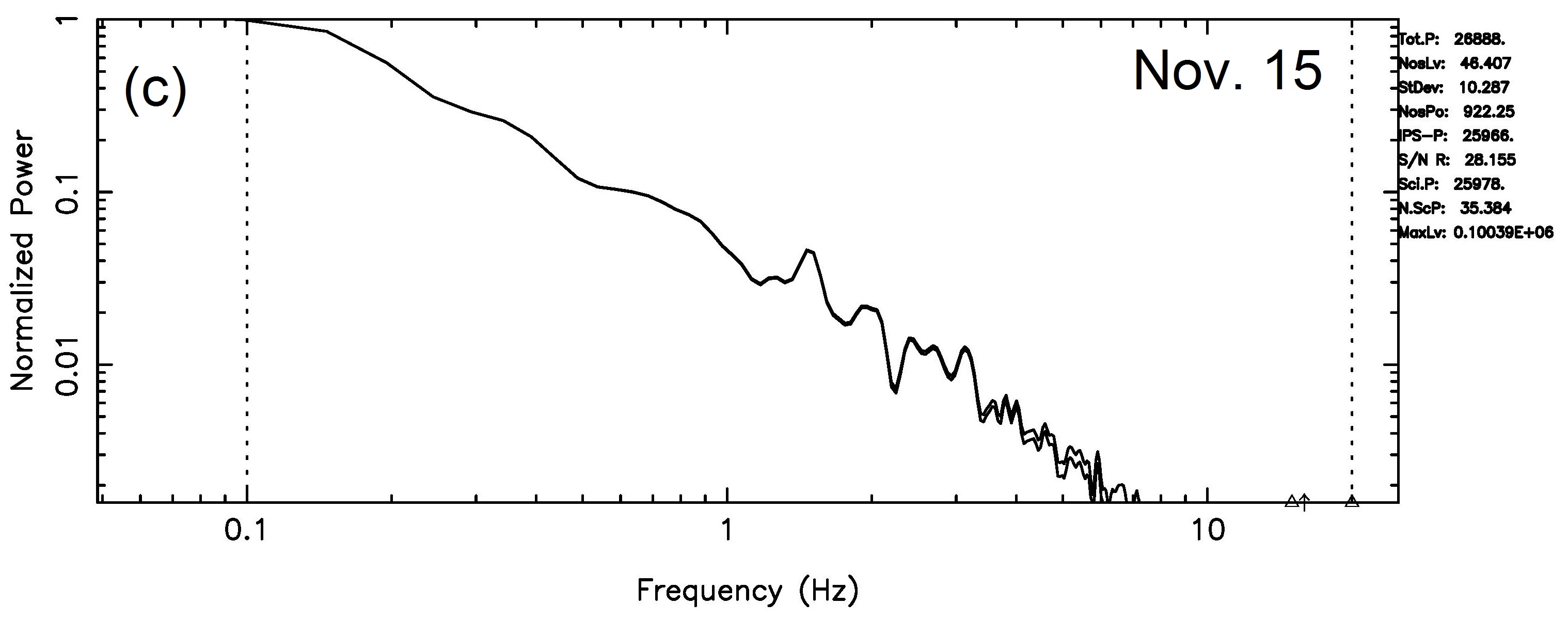}
                     \hspace*{0.00 \textwidth}
                     \includegraphics[width=0.50 \textwidth,clip=]{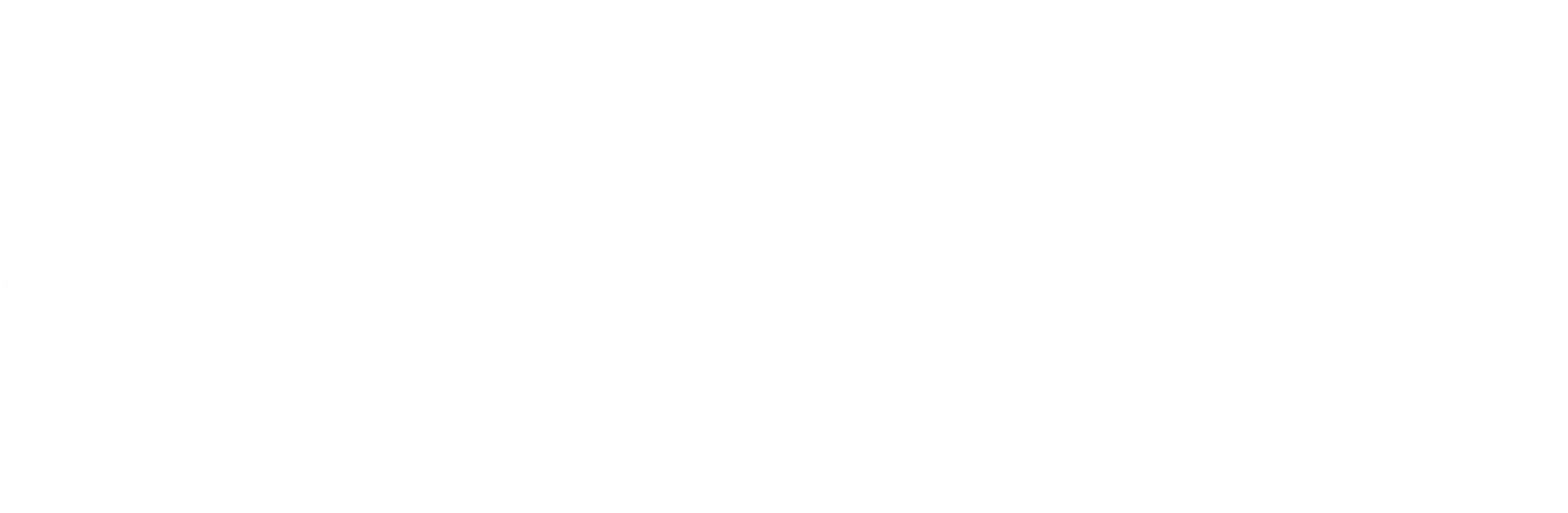}
                     }
           \vspace{0.00 \textwidth} 
      \caption{
      Power spectra of IPS for 1148$-$00 on (a) 13, (b) 14, (c) 15, (d) 16, and (e) 17, November 2013. 
      In each spectrum, the spectral power is normalized to the maximum value, and a pair of vertical dotted 
      lines indicates a range of the IPS component. 
      In panels (a) and (b), the horizontal broken line represents the noise level. 
      }
      \label{fig:fig5}
      \end{center}
      \end{figure}

\begin{figure}[!p]
      \begin{center}
      \centerline{\includegraphics[width=0.50 \textwidth,clip=]{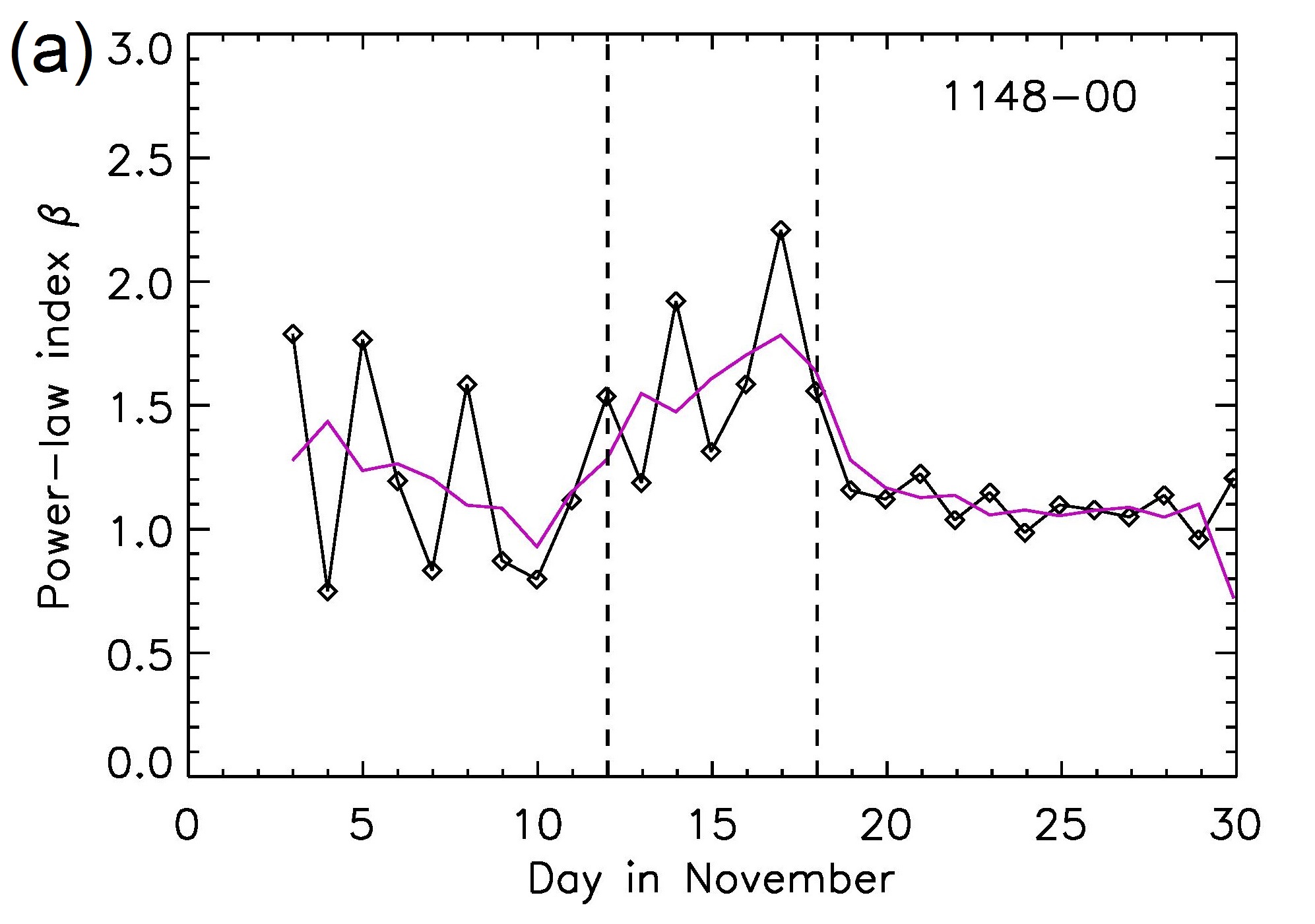}
                     \hspace*{0.01 \textwidth}
                     \includegraphics[width=0.50 \textwidth,clip=]{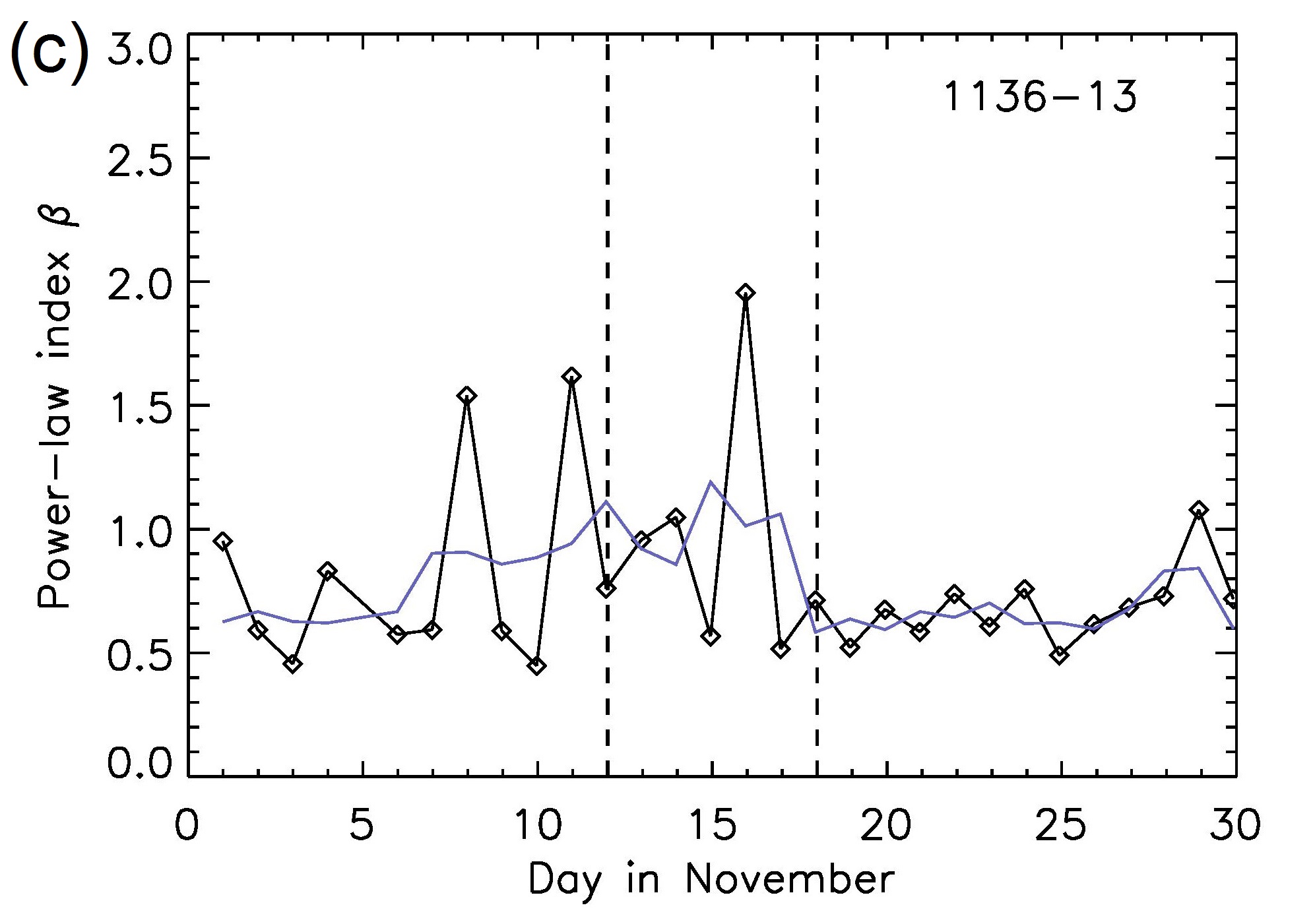}
                     }
           \vspace{0.010 \textwidth} 
         \centerline{\includegraphics[width=0.50 \textwidth,clip=]{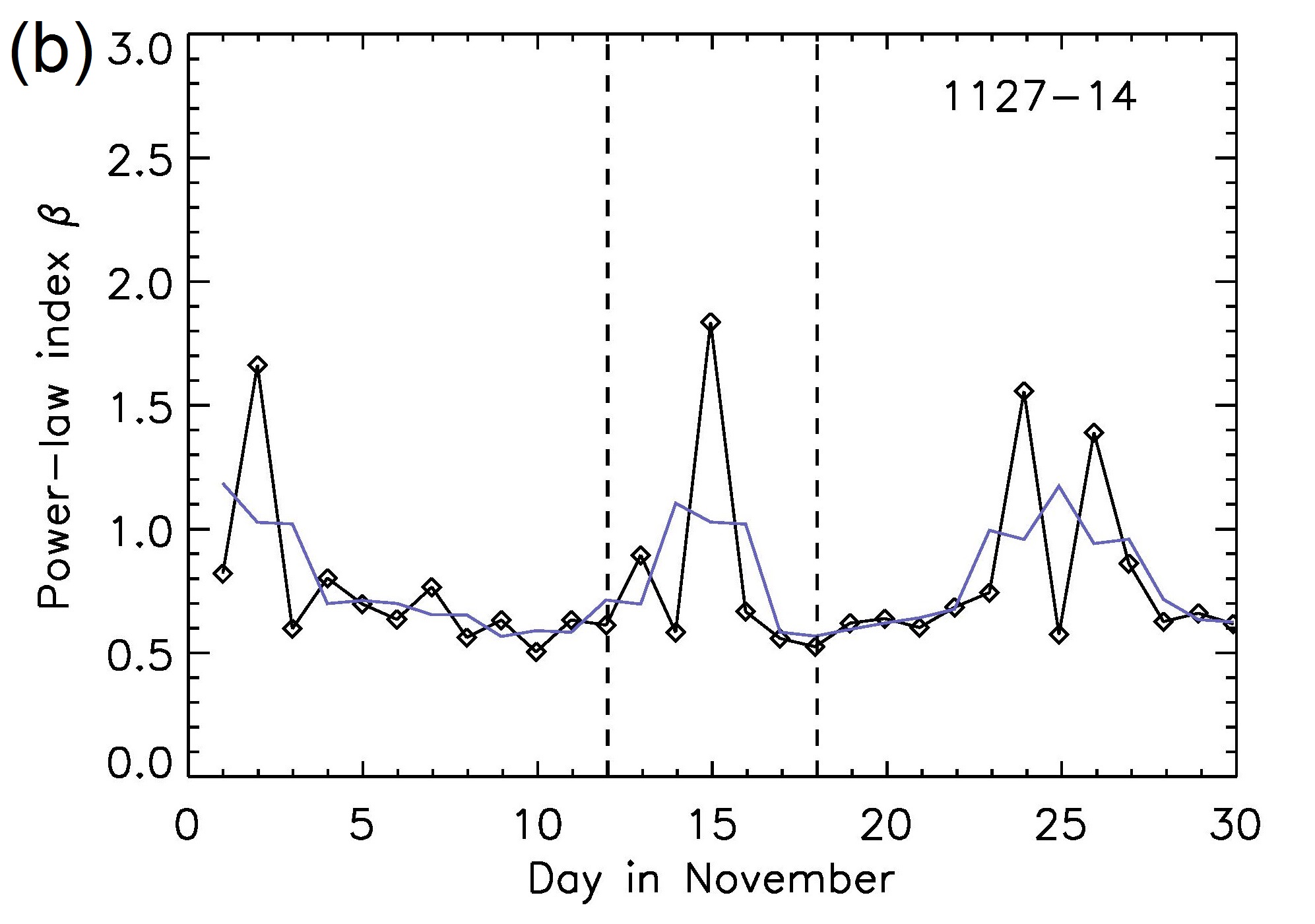}
                     \hspace*{0.01 \textwidth}
                     \includegraphics[width=0.50 \textwidth,clip=]{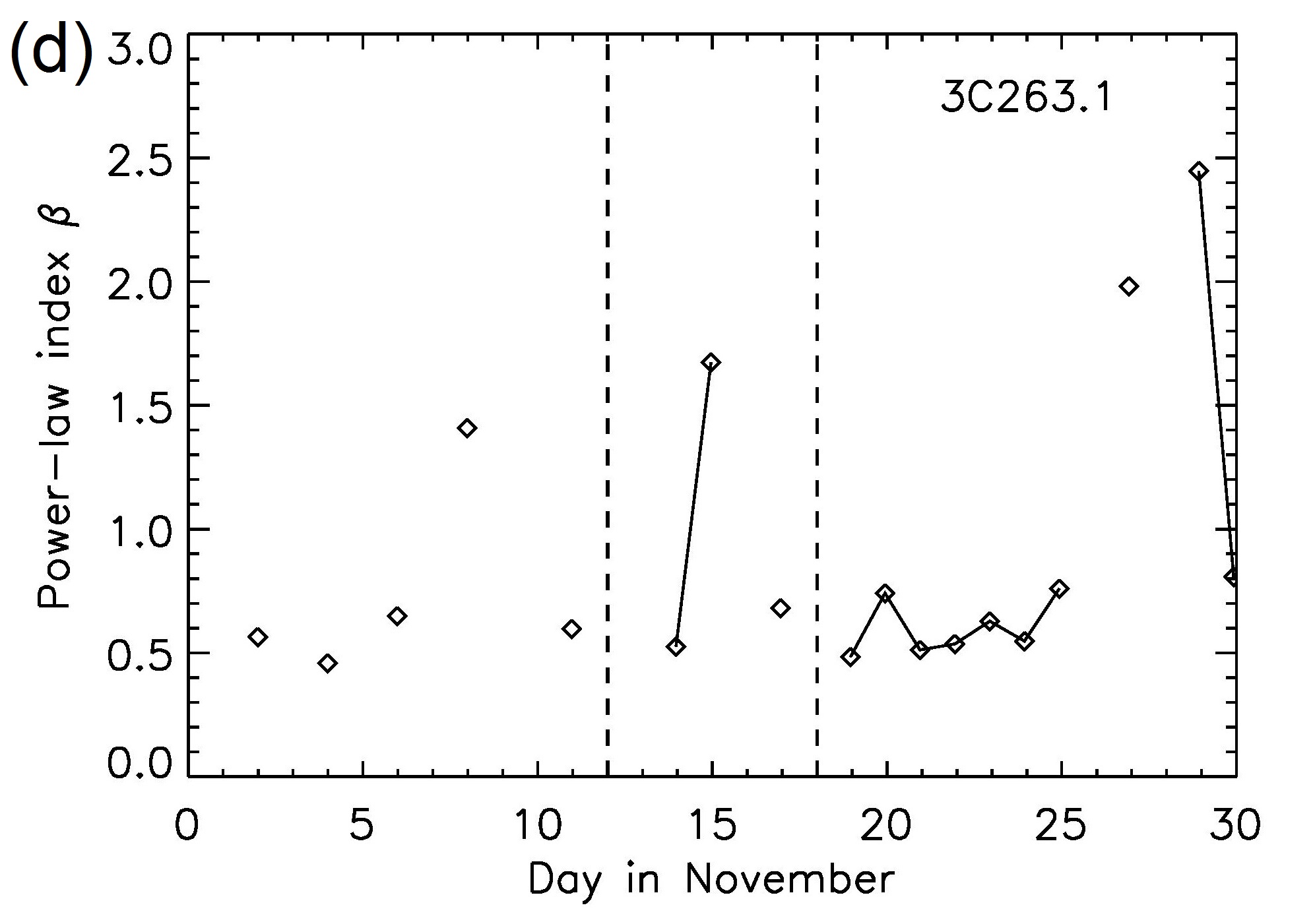}
                     }
           \vspace{0.00 \textwidth} 
      \caption{
      Daily variations of the power-law index ${\beta}$ for IPS power spectra of (a) 1148$-$00, (b)1127-14, (c)1136-13,
      and (d) 3C263.1 in November 2013. In each panel, diamonds denote data points, and successive ones are connected by solid lines. 
      The colored polyline shows a three-point moving average of ${\beta}$. 
      A pair of vertical broken lines indicates a period between the 12th and 18th when 1148$-$00 was occulted by 
      the ISON's plasma tail. 
      }
      \label{fig:fig6}
      \end{center}
      \end{figure}

We plot power spectra of radio-intensity fluctuation for 1148$-$00 during 
November 13\,--\,17 in Figure \ref{fig:fig5}. We mention that spectra of the 16 and 17 November IPS enhancements are similar to those 
of the 13 and 15 November events, respectively, and the radio fluctuation on 14th exhibits a higher noise level than the others 
for the spectrum. The slope of power spectra in the log-log graph can be fitted by the following power-law function:
\begin{equation}
    \label{wq.powerlaw}
P(f) = {\alpha}f^{-\beta}, 
\end{equation}
where $P(f)$ is the power spectral density in arbitrary units, $f$ is the frequency, and $\alpha$ and $\beta$  are the power-law coefficient 
and index, respectively, which are constants. We estimated the power-law index $\beta$ for 1148$-$00, 1127$-$14, 1136$-$13, and 3C263.1 to 
examine the change of IPS spectra by the cometary plasma tail and ICME. Figure \ref{fig:fig6} shows daily variations and three-days moving 
averages of $\beta$ for them in November 2013. We note that 18 of 30 power spectra for 3C263.1 can be fitted well by the power-law function, and 
intermittent data of $\beta$ for that are shown in the panel (d) in Figure \ref{fig:fig6}. 

\begin{figure}[!p]
      \begin{center}
      \centerline{\includegraphics[width=0.9\textwidth,clip=]{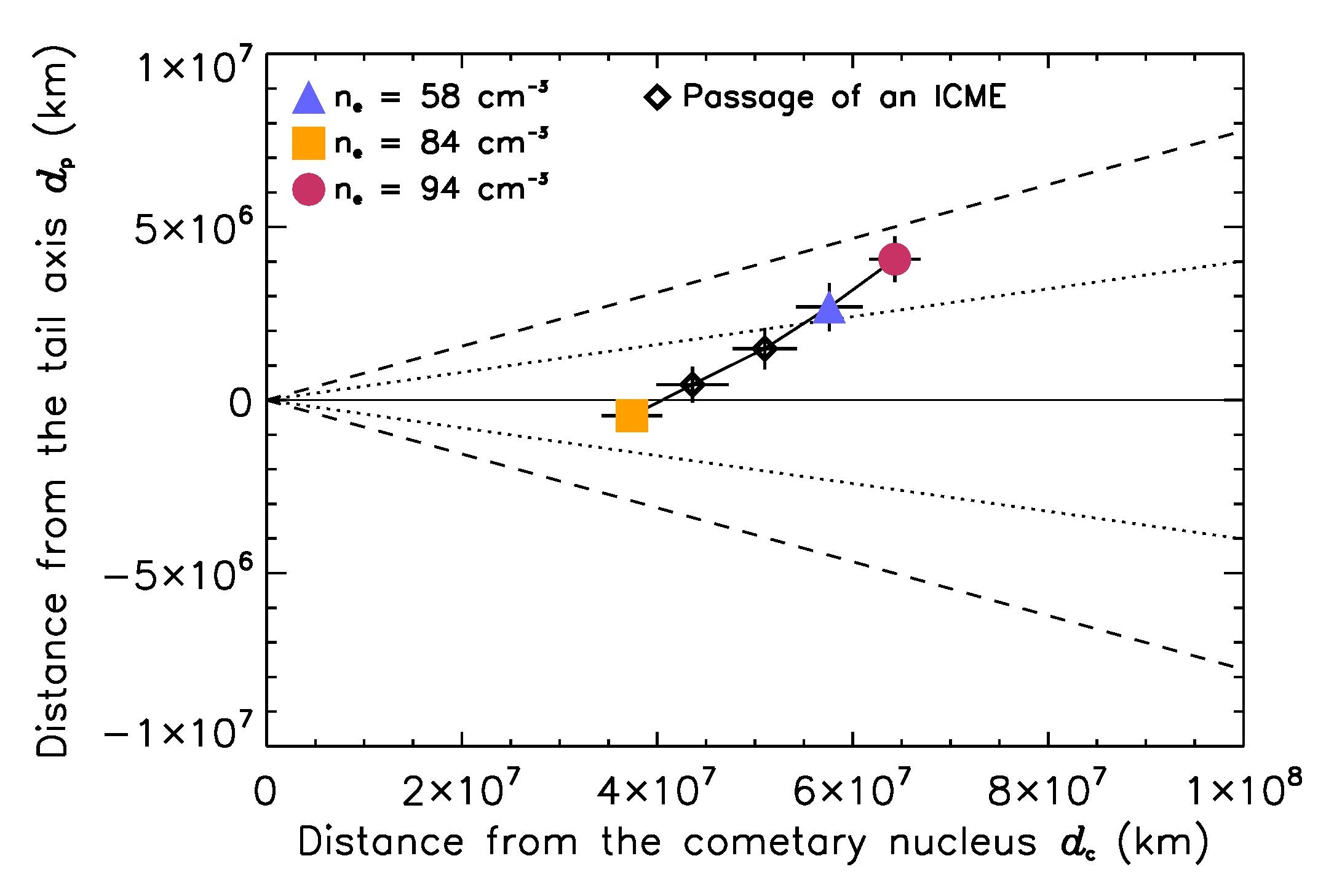}}
      \vspace{0.010 \textwidth}
      \caption{
      Electron densities of the ISON's plasma tail measured on 13, 16, and 17 November 2013. The line-of-sight from 1148$-$00 moves 
      from bottom left to top right. In the vertical axis, the negative number of distance corresponds to the south side of the plasma tail, 
      while the other is the north side. The filled square, triangle, and circle show positions of the line-of-sight on 13th, 16th, and 17th, 
      respectively, and their colors represent electron densities. Diamonds indicate positions of the line-of-sight on 14th and 15th in which 
      their densities do not appear because of ICME passing. 
      Symbols are connected by solid lines. The horizontal solid line denotes the cometary tail axis. Pairs of dotted 
      and broken lines represent boundaries of the plasma tail with the top angle of $\theta_\mathrm{tail} = 4.6$\degree and 8.9\degree, 
      respectively. 
      }
      \label{fig:fig7}
      \end{center}
      \end{figure}

\begin{figure}[!p]
      \begin{center}
      \centerline{\includegraphics[width=0.9\textwidth,clip=]{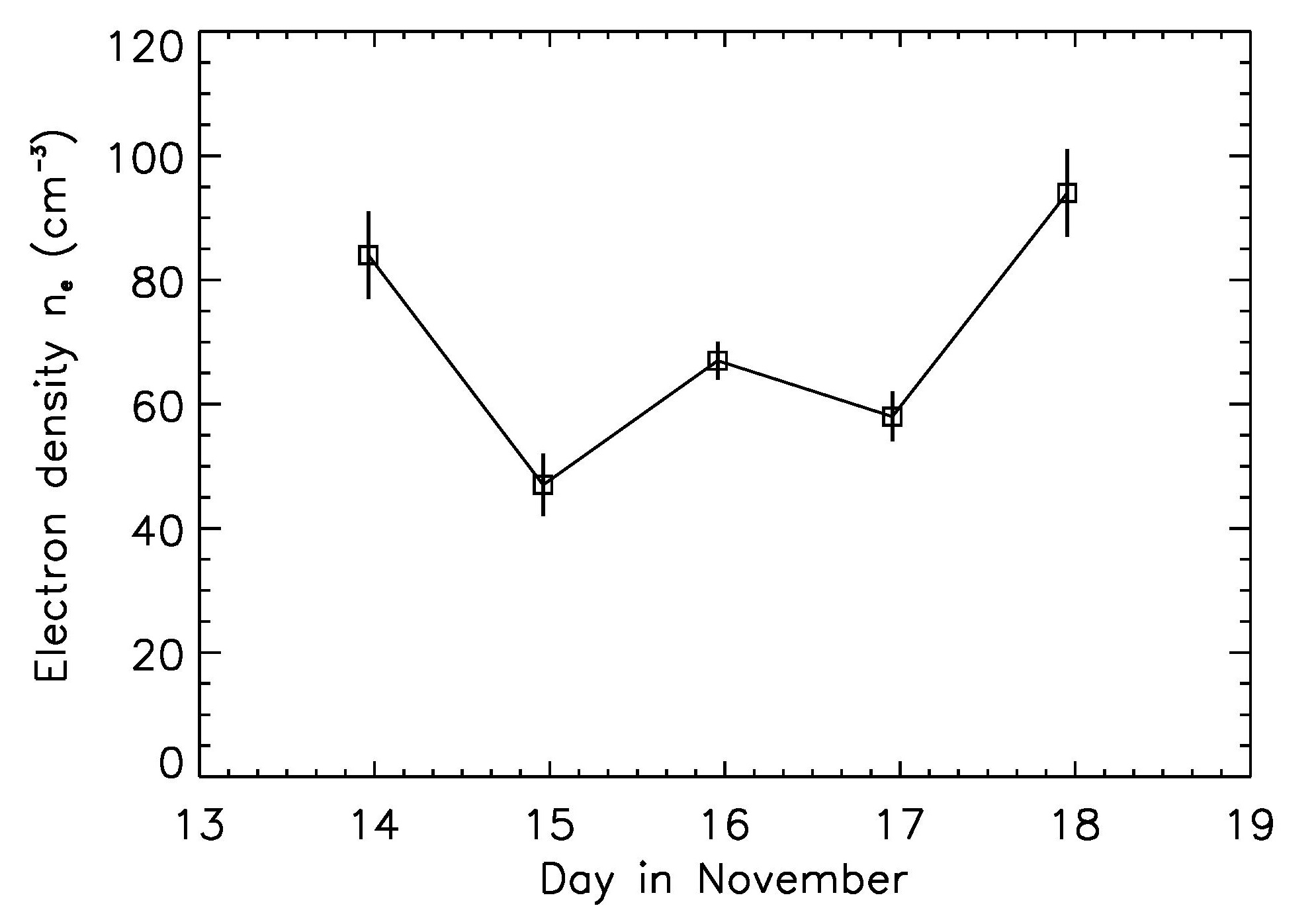}}
      \vspace{0.010 \textwidth}
      \caption{
      Time variation of electron density in the ISON's plasma tail between 13 and 17 November 2013. 
      }
      \label{fig:fig8}
      \end{center}
      \end{figure}

Although an ICME passed through the line-of-sight from 1148$-$00 on November 14 and 15, we calculated the electron density 
of plasma tail for all candidates. Table \ref{tab:electrondensity} gives ${d_\mathrm{c}}$, ${d_\mathrm{p}}$, and the average density 
${n_\mathrm{e}}$ with their errors for the five IPS enhancements. 
We emphasize that errors of ${d_\mathrm{c}}$ and ${d_\mathrm{p}}$ in this table are half values of the line-of-sight migration with respect to 
the ISON's nucleus in a day in the tail-axis and its perpendicular directions, respectively. 
We also note that errors of ${n_\mathrm{e}}$ are calculated by integrations of Equations (\ref{eq:gvalue2}) and (\ref{eq:cases}) with 
errors of ${W_\mathrm{tail}}$ which are deduced from those of ${d_\mathrm{c}}$ and ${d_\mathrm{p}}$, and they are minimal because 
other sources such as the density change of the ambient solar wind itself along the line-of-sight are neglected for the perturbation. 
Figure \ref{fig:fig7} shows the distribution of average electron density in the plasma tail of Comet ISON. In this figure, error bars represent 
errors of ${d_\mathrm{c}}$ and ${d_\mathrm{p}}$, which are listed in Table \ref{tab:electrondensity}. Because the IPS enhancements on the 14th 
and 15th are caused by both the ISON's plasma tail and ICME, values of ${n_\mathrm{e}}$ derived from them do not appear in Figure \ref{fig:fig7}. 
Figure \ref{fig:fig8} shows a daily variation of ${n_\mathrm{e}}$ deduced from scintillation enhancements of 1148$-$00. In this figure, 
vertical error bars represent errors of ${n_\mathrm{e}}$, and all data of ${n_\mathrm{e}}$ including the ICME event appear. 

      %
     
       %
        \begin{table}[!p]
        \caption{
        Electron densities in the plasma tail of Comet ISON deduced from IPS observations.
        }
        \label{tab:electrondensity}
        \scalebox{0.90}{
        \begin{tabular}{cccccc}
        \hline
Date & Time & $d_\mathrm{c}$ & $d_\mathrm{p}^\dagger$ & ${n_\mathrm{e}}$ & Note \\ 
~~~~ & (UT) & ($\times 10^{7}~\mathrm{km}$) & ($\times 10^{5}~\mathrm{km}$) & (${\mathrm{cm^{-3}}}$) & ~~~ \\
        \hline
Nov.~13 & 23:09 & $3.74 \pm 0.31$ & $-4.50 \pm 4.50$ & $84 \pm 7$ & ~~~ \\
Nov.~14 & 23:05 & $4.36 \pm 0.37$ & $~4.49 \pm 5.20$ & $47 \pm 5$ & An ICME passing \\
Nov.~15 & 23:01 & $5.10 \pm 0.33$ & $14.88 \pm 6.00$ & $67 \pm 3$ & An ICME passing \\
Nov.~16 & 22:57 & $5.76 \pm 0.34$ & $26.87 \pm 6.95$ & $58 \pm 4$ & ~~~ \\
Nov.~17 & 22:53 & $6.43 \pm 0.26$ & $40.77 \pm 6.55$ & $94 \pm 7$ & ~~~ \\
        \hline
\multicolumn{6}{l}{
$^\dagger$ ``$-$'' corresponds to the south side of the plasma tail.
} \\
        \end{tabular}
        }
        \end{table}
     

      %
     
      %
     
      %

\section{Discussion and conclusions}
  \label{discussion}

We present positive results of the 1148$-$00 occultation by the plasma tail of Comet ISON. The vantage point of STEREO/HI enables us to 
confirm the passage of an ICME during IPS observations and so select IPS enhancements due to the ISON's tail from our data. 
No large ICMEs pass through Comet ISON around 23:00 UT on November 13, 16, and 17, and so it is suggested that enhanced \textit{g}-values 
of 1148$-$00 on these days are probably due to the occultation by the ISON's plasma tail. From Figure \ref{fig:fig1}, we find that the distance 
between the ISON's nucleus and the line-of-sight from 1148$-$00 is larger than the maximum length of the ISON's tail listed in Table \ref{tab:isontail}, 
although the line-of-sight is inside the prolonged tail (see Figure \ref{fig:fig7}) during the 16\,--\,18th. We consider that the ISON's plasma tail is 
extended more than $4.47 \times 10^{7}~\mathrm{km}$, and consists of the visible tail up to and an invisible tail beyond this. 
We argue the validity of this hypothesis because the plasma tail of Comet Hyakutake was detected at the distance of $5.5 \times 10^{8}~\mathrm{km}$ 
from the cometary nucleus by in situ measurement (\citealp{Jones2000}) while its visible part had a maximum angular length of 70\degree 
(corresponding to $\approx 5 \times 10^{7}~\mathrm{km}$) (\citealp{James1998}; see also \url{http://epod.usra.edu/library/comet_011510.html}). 

In earlier studies, \cite{Janardhan1992} reported a systematic slope change of IPS spectra for a radio source 3C459 during the passage of 
Comet Halley's tail. \cite{Roy2007} also reported that power spectra of B0019$-$000 exhibited an intensity excess of 
its lower frequency part at the closest approach of Comet Schwassmann-Wachmann 3-B. Now we examine these phenomena from our observations. 
Figure \ref{fig:fig5} shows the power spectra of 1148$-$00 during November 13\,--\,17, 2013. From this figure, we find that two types of spectra 
appear alternately during the 13\,--\,17th except for the 14th and these spectra do not largely different from normal ones. 
From Figure \ref{fig:fig6}, we find that the power spectra of 1148$-$00 have an increase in the power-law index $\beta$ during 
the occultation by the ISON's plasma tail. A moving average of $\beta$ for 1148$-$00 keeps $\approx1.6$ during November 12\,--\,18 while 
$\approx1.1$ before the 12th and after the 18th. This variation is probably attributed to the ISON's tail. 
The larger $\beta$ means the steeper slope of power spectra and that the turbulence with larger spatial scales is predominant in the plasma. 
Therefore, this result suggests that the cometary plasma tail has different plasma properties from the non-disturbed solar wind. 
Other sources 1127$-$14, 3C263.1, and 1136$-$13 show a short lasted increase of $\beta$ on November 14 and 15, respectively, which are related 
to the passage of an ICME. The degree of $\beta$ increasing for them is comparable to that for 1148$-$00. 
We emphasize that radio sources often show their deformed IPS spectra, which are caused by interplanetary disturbances such as ICMEs, 
in our routine observations. 

      %
      \begin{figure}[!p]
      \begin{center}
      \centerline{\includegraphics[width=0.9\textwidth,clip=]{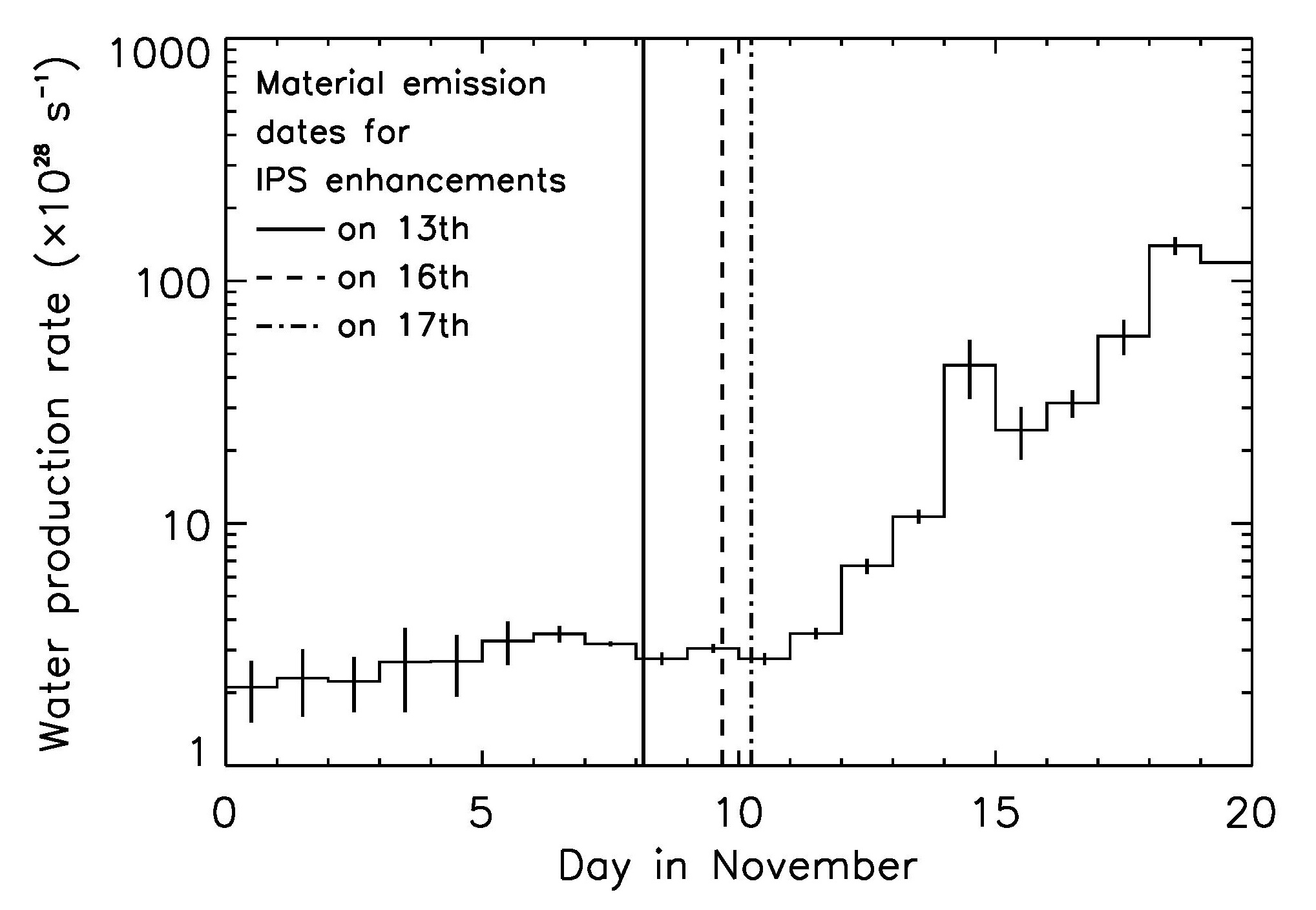}}
      \vspace{0.010 \textwidth}
      \caption{
      Comparison between the water production rate of Comet ISON and a set of material emission dates for the IPS enhancements on 
      November 13, 16, and 17, 2013. The histogram with error bars shows the daily averaged water production rate 
      (in units of $10^{28}~\mathrm{s^{-1}}$) as a function of time, which was studied by \cite{Combi2014}. The solid, broken, and 
      dash-dotted lines indicate the expected dates of material emissions from the ISON's nucleus for the IPS enhancements of 1148$-$00 
      on 13th, 16th, and 17th, respectively. These dates are calculated using a velocity equation of plasma blobs flowing in the tail 
      deduced from Comet Halley (\citealp{Celnik1987}). 
      }
      \label{fig:fig9}
      \end{center}
      \end{figure}
     
From Table \ref{tab:electrondensity} and Figure \ref{fig:fig8}, the passage of an ICME does not seem to significantly affect 
the estimation of the electron density in the ISON's plasma tail, although we exclude the 14 and 15 November IPS enhancements from consideration 
in Figure \ref{fig:fig5} because of the ICME. 
\cite{Meyer-Vernet1986} derived the electron density in a cross section of the plasma tail from in situ observations of Comet Giacobini-Zinner. 
They showed that the electron density became the maximum value on the tail axis and decreased with an increase in distance from 
that. From Figure \ref{fig:fig5}, electron densities of the ISON's tail derived from the 13 and 16 November IPS enhancements 
are consistent with the earlier study. On the other hand, we find an unexpected increase of the electron density at a point close to the outermost 
boundary on November 17 from the same figure. To discuss the cause of this variation, we examine the outgassing activity of 
Comet ISON. 

\cite{Combi2014} determined the water production rate of Comet ISON from October 24 to November 24 using data of 
the \textit{Solar Wind Anisotropies} (SWAN) instrument on SOHO. According to them, the water ejection from the ISON's nucleus increased rapidly by 
two orders of magnitude (from $10^{28}$ up to $10^{30}$ molecules per second) during November 12\,--\,18. On the other hand, 
many web pages indicated that Comet ISON became bright suddenly and visible to the naked eye on the 14th 
(e.g. \url{www.space.com/23591-comet-ison-visibility-naked-eye.html}). A combination of these facts suggests that an outburst of Comet ISON begins 
about two days earlier than a brightening thereof. Now we calculate dates for material emissions from the ISON's nucleus which caused 
the above IPS enhancements in the plasma tail. However, we do not yet know the velocity distribution of plasma flowing in the ISON's tail. 
Instead, we use the velocity equation of plasma condensations in the cometary tail deduced from another comet. \cite{Celnik1987} 
analyzed the dynamics of Comet Halley's plasma tail up to the distance of $5.0 \times 10^{7}~\mathrm{km}$ from its nucleus using photographical 
observations. They derived the following equation:
\begin{equation}
 \label{eq:velocity}
v(t) = 2.77 \times 10^{-4}t + 4.5,
\end{equation}
where $t$ is the time, assuming the constant acceleration in the tail. We use this equation as a representative description of plasma velocity in 
the cometary tail and determine the material emission dates from its integration. Figure \ref{fig:fig9} shows a comparison between the ISON's water production 
rate and a set of material emission dates for the IPS enhancements on November 13, 16, and 17, 2013. From this figure, we find that substances which caused 
these IPS enhancements are ejected from the ISON's nucleus before the outburst. Even if a parcel of ionized molecules is assumed to reach 
the line-of-sight from 1148$-$00 two days after its ejection on the basis of the above suggestion, the material emission date for 
the 13 November IPS enhancement is just before the beginning of the outburst. Hence, we conclude that at least the IPS enhancement of 1148$-$00 on 
the 13th probably does not stem from the outburst of Comet ISON. The remaining IPS enhancements are observed at the rising phase of the ISON's 
water production rate, while their associated gas ejections are expected to occur just before the beginning of the outburst as shown in Figure \ref{fig:fig9}. 
To discuss precisely whether the ISON's outburst relates to them, we need to know the velocity distribution of the ISON's plasma tail before 
its perihelion because the plasma velocity becomes the minimum on the tail axis and increases to the speed of the solar wind with the distance from that 
(\citealp{Bame1986}; \citealp{Neugebauer2007}). 
The measurement points for the 16 and 17 November IPS enhancements locate in the sparse region of the plasma tail as shown in Figure \ref{fig:fig7}. 
We expect that the solar wind merges with the cometary plasma, and the flow of mixed plasma is highly turbulent in this region. 
There is another possibility that the tail disconnection event is responsible for the electron density increasing on the 17th. The variation of 
the solar wind, e.g. an increase in the dynamic pressure or a change of the interplanetary magnetic field polarity, may causes the tail disconnection 
event (\citealp{Voelzke2005}). \cite{Vourlidas2007} reported that an ICME disconnected and carried off Comet Encke's tail. 
If the disconnected plasma traveled from the ISON's coma at the speed of the solar wind ($\approx450~\mathrm{km~s^{-1}}$), it would traverse 
the line-of-sight from 1148$-$00 on the 17th at 22:53 UT within approximately 40 h from its departure. From this assumption, we estimate 
that a disconnection event should occur by 7:00 UT on the 16th at the latest. 

We obtained a rare opportunity to investigate the plasma tail of Comet ISON using the IPS and STEREO/HI observations. In this study, we identified 
IPS enhancements of a radio source 1148$-$00 on November 13, 16, and 17, which were probably caused by the occultation of the ISON's plasma tail. 
Our examinations of them revealed that a change of the IPS power spectra was observed during the passage of 
Comet ISON. We estimated the electron density of the plasma tail. 
Although we cannot observe Comet ISON again, a combination of the ground-based IPS and space-borne interplanetary imager observations provides 
a useful means to study the plasma tail for other various comets. 



\section*{Acknowledgments}

The IPS observations were carried out under the solar wind program of the Solar-Terrestrial Environment Laboratory (STEL), 
Nagoya University. We thank the STEREO Science Center for use of their web service and STEREO/HI data. STEREO is 
the third mission in the Solar-Terrestrial Probes program by the National Aeronautics and Space Administration (NASA). 
We acknowledge use of the SOHO/LASCO CME catalog; this CME catalog is generated and maintained at the CDAW Data Center 
by NASA and the Catholic University of America in cooperation with the Naval Research Laboratory. SOHO is a project of 
international cooperation between the European Space Agency and NASA. 
We thank the Solar Software Library for use of the STEREO analysis software. 
We acknowledge G. Rhemann and M. J{\"{a}}ger for use of their images of Comet ISON. 
This work was supported by the IUGONET Project of MEXT, Japan.

\section*{References}

  \bibliographystyle{elsarticle-harv} 
  \bibliography{references}







\end{document}